\documentclass[11pt,a4paper]{article}

\usepackage[T1]{fontenc}
\usepackage{lmodern,microtype}
\usepackage[margin=27mm]{geometry}

\usepackage{amsmath,amssymb,amsthm,mathtools,mathrsfs}
\usepackage{booktabs,tabularx,array,enumitem}
\usepackage{needspace}

\usepackage{xcolor}
\usepackage[
    colorlinks=true,
    linkcolor=blue!45!black,
    citecolor=blue!45!black,
    urlcolor=blue!45!black
]{hyperref}

\usepackage{tikz}
\usetikzlibrary{arrows.meta}

\usepackage{fancyhdr}

\allowdisplaybreaks[2]
\numberwithin{equation}{section}

\newtheorem{theorem}{Theorem}[section]
\newtheorem{lemma}[theorem]{Lemma}
\newtheorem{proposition}[theorem]{Proposition}
\newtheorem{corollary}[theorem]{Corollary}

\theoremstyle{definition}
\newtheorem{example}[theorem]{Example}

\newtheorem{remark}[theorem]{Remark}

\newcommand{\R}{\mathbb R}
\newcommand{\C}{\mathbb C}
\newcommand{\N}{\mathbb N}

\newcommand{\cH}{\mathcal H}
\newcommand{\cB}{\mathcal B}
\newcommand{\cF}{\mathcal F}

\newcommand{\id}{\mathrm{id}}
\newcommand{\ii}{\mathrm i}

\newcommand{\norm}[1]{\lVert#1\rVert}
\newcommand{\ip}[2]{\langle#1,#2\rangle}

\DeclareMathOperator{\Tr}{Tr}
\DeclareMathOperator{\supp}{supp}
\DeclareMathOperator{\Ran}{Ran}
\DeclareMathOperator{\Dom}{Dom}
\DeclareMathOperator{\Rea}{Re}
\DeclareMathOperator{\Ima}{Im}
\DeclareMathOperator{\rad}{rad}

\newcommand{\HS}{\mathrm{HS}}

\title{\textbf{A Conditional Quantum Kac--Bernstein Theorem}}

\author{%
Hesam Hosseini\textsuperscript{1},
Jinpei Zhao\textsuperscript{1}, and
Masahito Hayashi\textsuperscript{2,3,4}\\[7pt]
{\footnotesize
\begin{tabular}{>{\centering\arraybackslash}p{0.85\textwidth}}
\textsuperscript{1}Department of Information Engineering,
The Chinese University of Hong Kong, Hong Kong\\[3pt]
\textsuperscript{2}School of Data Science,
The Chinese University of Hong Kong, Shenzhen,
Longgang District, Shenzhen 518172, China\\[3pt]
\textsuperscript{3}International Quantum Academy,
Futian District, Shenzhen 518048, China\\[3pt]
\textsuperscript{4}Graduate School of Mathematics,
Nagoya University, Nagoya 464-8602, Japan
\end{tabular}}
}

\date{}

\begin{document}

\maketitle

\begin{abstract}

We establish a conditional quantum Kac–Bernstein theorem for mixing of two independent, finite-mode bosonic inputs with quantum memories.  We characterize exactly when the outputs are conditionally independent given both memories. Each input then decomposes into orthogonal memory sectors sharing a common Gaussian state on the physical system and a bosonic memory factor, up to sector-dependent physical displacements.  In particular, triviality of the common bosonic factor implies separability of the inputs, as occurs whenever either memory has finite-dimensional support. 
\end{abstract}

\section{Introduction}
\label{sec:overview}

The Kac--Bernstein theorem characterizes Gaussian laws
through the independence of the sum and difference of
two independent random vectors. For independent random
vectors $X_1,X_2$, let
$\varphi_i(s):=\mathbb E[e^{\ii s^{\mathsf T}X_i}]$
denote their characteristic functions.
Independence of the sum and difference is equivalent
to the Bernstein equation
\begin{equation}\label{eq:classicalbernstein}
\varphi_1(s+t)\varphi_2(s-t)
=\varphi_1(s)\varphi_2(s)
\varphi_1(t)\varphi_2(-t).
\end{equation}
Its solutions are Gaussian characteristic functions
with a common covariance, although their means may
differ \cite{Kac,Bernstein}. In bosonic quantum systems,
a balanced beam splitter implements the corresponding
sum--difference transformation. Independence of its
two outputs similarly characterizes Gaussian inputs
with a common covariance in the absence of quantum
memory \cite{Springer,Cuesta}.

In classical information theory, the independence
of sum--difference transformations often arises only
after conditioning on auxiliary random variables.
Conditional Gaussian characterizations play an
important role in proofs of Gaussian optimality
for broadcast channels, entropy inequalities, and
relay channels
\cite{GengNair,AnantharamJogNair,ElGamalGohariNair}.
These arguments use conditional independence of
rotated copies to establish Gaussian conditional
laws with a common covariance. Their means may
still depend on the auxiliary variables, so the
unconditional distributions need not be Gaussian.
The conditioning is therefore essential.

To illustrate this point, consider two independent
classical input--memory pairs $(X_1,U_1)$ and
$(X_2,U_2)$. If the sum and difference of $X_1$
and $X_2$ are conditionally independent given
$(U_1,U_2)$, the classical Kac--Bernstein theorem
applies to the conditional input laws. Their
characteristic functions then take the form
\begin{equation}\label{eq:introclassical}
\varphi_i(s\mid u_i)
=\exp\left(
\ii b_i(u_i)^{\mathsf T}s
-\tfrac12s^{\mathsf T}\Sigma s
\right),\qquad i=1,2,
\end{equation}
where the covariance $\Sigma$ is common to both
inputs and independent of the memory values,
while the conditional means $b_i(u_i)$ may differ.

Extending these classical Gaussian optimality
arguments to quantum side information calls for
a corresponding conditional characterization.
The classical conditioning argument, however,
is no longer available: quantum memory observables
need not commute, and physical--memory correlations
cannot generally be described by conditioning
on simultaneous memory values. Quantum conditional
independence is instead connected with the structure
of quantum Markov states
\cite{HJPW,JPP}. This leads to the question of
what conditional independence of sum--difference
outputs implies for bosonic inputs accompanied
by quantum memories.

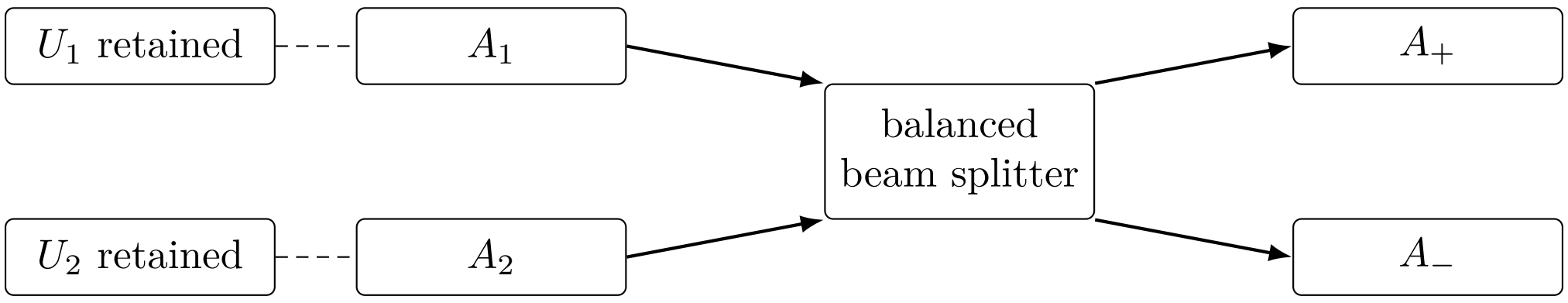
\begin{figure}[htbp]
\centering
\begin{tikzpicture}[
box/.style={draw,rounded corners=2pt,
minimum width=2.3cm,minimum height=.65cm,
align=center,font=\small},
arr/.style={-{Latex[length=2mm]},thick}]

\node[box] (a1) at (-4,.9) {$A_1$};
\node[box] (a2) at (-4,-.9) {$A_2$};

\node[box] (u1) at (-7,.9) {$U_1$ retained};
\node[box] (u2) at (-7,-.9) {$U_2$ retained};

\node[box,minimum height=1.15cm]
(bs) at (0,0) {balanced\\beam splitter};

\node[box] (ap) at (4,.9) {$A_+$};
\node[box] (am) at (4,-.9) {$A_-$};

\draw[arr] (a1.east)--(bs.north west);
\draw[arr] (a2.east)--(bs.south west);
\draw[arr] (bs.north east)--(ap.west);
\draw[arr] (bs.south east)--(am.west);

\draw[densely dashed] (u1.east)--(a1.west);
\draw[densely dashed] (u2.east)--(a2.west);

\end{tikzpicture}
\caption{Balanced mixing of two independent
inputs $\rho_1$ and $\rho_2$. The beam splitter
acts only on the physical systems, while
$U_1U_2$ is retained as the conditioning memory.}
\label{fig:experiment}
\end{figure}

We address this question for two  finite-mode bosonic systems $A_1,A_2$, each accompanied by a quantum memory $U_1,U_2$ on
a separable Hilbert space. The normal input
states $\rho_1$ and $\rho_2$, with
$\rho_i=\rho^{(i)}_{A_iU_i}$, are independent
but need not be identical. A balanced beam
splitter acts only on the physical systems,
transforming their quadrature vectors
$R_1,R_2$ into $(R_1+R_2)/\sqrt2$ and
$(R_1-R_2)/\sqrt2$; see
Figure~\ref{fig:experiment} and
\eqref{eq:beam}. For the resulting state
$\omega$, we characterize exactly when
\begin{equation}\label{eq:openingCMI}
I(A_+:A_-\mid U_1U_2)_\omega=0.
\end{equation}
We assume finite physical energies and finite
memory entropies, without requiring Gaussian
inputs or finite-dimensional memories.
Bosonic mixing with quantum side information
also arises in conditional entropy power
inequalities \cite{DePalmaTrevisan}. Here,
our aim is to determine the input states
satisfying the exact conditional independence
condition \eqref{eq:openingCMI}.

To derive the quantum Bernstein equation, we use
the \emph{partial characteristic functions}
of the inputs. Let $W_{A_i}(s)$ denote the
physical Weyl operators \cite{Weedbrook}, and set
\begin{equation}\label{eq:introoperator}
\Phi_i(s):=
\Tr_{A_i}[\rho_i(W_{A_i}(s)\otimes I)]
=\mu_i^{1/2}K_i(s)\mu_i^{1/2},
\qquad
\mu_i:=\Tr_{A_i}\rho_i.
\end{equation}
The normalized operators $K_i(s)$ are bounded
on the support of $\mu_i$. They allow us to
express conditional independence as an
operator-valued analogue of the classical
Bernstein equation.

We first show that the conditional independence
condition \eqref{eq:openingCMI} implies
\begin{equation}\label{eq:introbernstein}
K_1(s+t)\otimes K_2(s-t)
=K_1(s)K_1(t)\otimes
K_2(s)K_2(-t),
\end{equation}
the quantum analogue of
\eqref{eq:classicalbernstein}. Solving this
paired operator equation gives
\begin{equation}\label{eq:introclassification}
K_i(s)=e^{-s^{\mathsf T}Gs/4}V_i(s),
\qquad
V_i(s)V_i(t)
=e^{-\ii s^{\mathsf T}\Theta t/2}
V_i(s+t).
\end{equation}
The positive semidefinite matrix $G$ and
antisymmetric matrix $\Theta$ are common to
both inputs, although the unitary families
$V_i$ may differ.

To understand these unitary families, we study
their evolution under the modular groups of
the memory states. The resulting commutation
relations separate the memory fields into fixed
and rotating components. The rotating fields
satisfy canonical commutation relations, and
their frequencies determine the corresponding
thermal parameters. The Stone--von Neumann
theorem identifies finite collections of these
fields with standard bosonic Fock representations.
For the full countable family, we establish
that the occupation-number sectors exhaust
the original memory Hilbert spaces. This
allows us to identify a common thermal Fock
factor and separate it from the remaining
memory degrees of freedom.

With this structure in place, we obtain our
main result: a necessary and sufficient
characterization of \eqref{eq:openingCMI}.
Theorem~\ref{thm:main} gives the decompositions
\begin{equation}\label{eq:overviewnormal}
\begin{aligned}
\cH_{U_i}
&\simeq\cH_Q\otimes
\left(\bigoplus_{j\in\mathcal J_i}
\cH_{K_j^{(i)}}\right),\quad
\rho_i
\simeq\bigoplus_{j\in\mathcal J_i}
p_j^{(i)}\gamma_{AQ}^{b_j^{(i)}}
\otimes\nu_j^{(i)},
\qquad i=1,2.
\end{aligned}
\end{equation}
Here $\gamma_{AQ}$ is a common centered
Gaussian state on the physical system and
a bosonic memory factor $Q$, whose marginal
on $Q$ is a product of thermal states.
Each $\gamma_{AQ}^{b}$ is obtained by
displacing only the physical system.
The two inputs therefore share the same
Gaussian core, while their displacement
distributions, residual memory states, and
memory Hilbert spaces may differ.
Conversely, every pair of states of this
form satisfies \eqref{eq:openingCMI}.

The classification also restricts the
entanglement compatible with conditional
independence. If $Q$ is trivial, both inputs
are separable mixtures of displaced Gaussian
states correlated with orthogonal memory
subspaces (Corollary~\ref{cor:trivial-Q}).
This remains true for infinite-dimensional
memories. In particular, finite-dimensional
support of either memory forces $Q$ to be
trivial. Thus entangled equality states
require infinite-dimensional support for
both memories, and any physical--memory
entanglement is contained in the common
Gaussian factor.

\textbf{Organization of the paper.} 
Section~\ref{sec:preliminaries} introduces the bosonic
conventions, the mixing experiment, and the
normalized partial characteristic functions.
Section~3 states the main theorem and its
consequences, including the separability result
for trivial bosonic memory factors.
Section~\ref{sec:bernstein} derives the operator
Bernstein equation and modular commutation
relations and establishes the paired Bernstein
classification.
Section~\ref{sec:geometry} develops the modular
covariance-space framework and identifies the
common rotating memory geometry.
Section~\ref{sec:thermal} establishes the thermal
Fock-space realization of the common bosonic
memory factor, including the countably infinite
case.
Section~\ref{sec:reconstruction} reconstructs
the Gaussian normal form and proves the converse
implication.
The concluding section summarizes the main
findings.
The appendices provide technical proofs
concerning the Markov decomposition, modular
fields and their domains, occupation-number
sectors, and the thermal sandwich identity.

\section{Preliminaries and Definitions}

\label{sec:preliminaries}

\subsection{Bosonic conventions}
\label{sec:setup}
All Hilbert spaces are separable, and all states are normal. We identify
a normal state with its density operator. Let $A$ be an $n$-mode bosonic system, with $n<\infty$, and set $d=2n$. Write $R_A=(Q_1,P_1,\ldots,Q_n,P_n)^{\mathsf T}$ for its quadrature operators. The Weyl operators satisfy
\begin{equation}\label{eq:weylA}
 W_A(s)=e^{\ii s^{\mathsf T}R_A},\qquad
 W_A(s)W_A(t)
 =e^{-\ii s^{\mathsf T}\Omega_A t/2}W_A(s+t),
 \qquad
 \Omega_A=\bigoplus_{k=1}^n
 \begin{pmatrix}0&1\\-1&0\end{pmatrix}.
\end{equation}

For a state $\rho_A$, its scalar quantum characteristic function
is defined by $
\chi_{\rho_A}(s):=\Tr[\rho_AW_A(s)].
$ This Weyl-characteristic-function convention is standard in
continuous-variable quantum information \cite{Weedbrook}. The state is Gaussian if its characteristic function has the form
\begin{equation}\label{eq:gaussianchar}
\chi_{\Gamma,b}(s)
=\exp\!\left(-\frac14s^{\mathsf T}\Gamma s
+\ii b^{\mathsf T}s\right),
\end{equation}
where $b\in\mathbb R^d$ is the mean vector and $\Gamma$ is the covariance matrix. In this convention, the vacuum covariance is $I$, and $\Gamma$ is physical if and only if $\Gamma+\ii\Omega_A\succeq0$.

Physical displacements and the number Hamiltonian are given by
\begin{equation}\label{eq:displacement}
 D_A(b):=W_A(\Omega_A b),\qquad
 D_A(b)^*W_A(s)D_A(b)=e^{\ii b^{\mathsf T}s}W_A(s),\qquad
 H_A:=\tfrac12(R_A^{\mathsf T}R_A-nI).
\end{equation}

We denote by $\cF(\mathfrak h)$ the symmetric Fock space
over a complex Hilbert space $\mathfrak h$. For a finite
or countable index set $\mathcal I$,
$\cF(\ell^2(\mathcal I;\C))$ is the bosonic Fock space
with modes indexed by $\mathcal I$. We also write
$\ell^2(\mathcal I;\R^2)$ for the space of sequences
$f=(f_\ell)_{\ell\in\mathcal I}$, where
$f_\ell=(f_{\ell,1},f_{\ell,2})\in\R^2$ and
$\sum_\ell|f_\ell|^2<\infty$.

For $0<q<1$, the one-mode thermal state is
$\tau(q):=(1-q)\sum_{m=0}^\infty
q^m|m\rangle\langle m|$,
where $\{|m\rangle\}_{m\ge0}$ is the number-state basis.

A normal state on $A$ with countably many memory modes
is called Gaussian if its restriction to $A$ and
every finite subset of memory modes is Gaussian.

\subsection{The mixing experiment}

Let $A_1$ and $A_2$ be two $n$-mode bosonic systems,
each described by the same Hilbert space $\mathcal H_A$,
and let $U_1$ and $U_2$ be quantum memories
with possibly different separable Hilbert spaces
$\mathcal H_{U_1}$ and $\mathcal H_{U_2}$. Write $\rho_i$ for the state on $A_iU_i$ and set

$\mu_i:=\Tr_{A_i}\rho_i$. When discussing one input alone,
we omit the index $i$ when no confusion is possible. We assume throughout that

\begin{equation}\label{eq:assumptions}
 \Tr(\rho_{i,A_i}H_{A_i})<\infty,\qquad S(\mu_i)<\infty,
 \qquad i=1,2.
\end{equation}
Each $\cH_{U_i}$ is restricted to $\supp\mu_i$, so $\mu_i$ is faithful
there. Independence means that the joint input is $\rho_1\otimes\rho_2$.
The output state is

\begin{equation}\label{eq:output}
 \omega:=(B_A\otimes I_{U_1U_2})(\rho_1\otimes\rho_2)
 (B_A^*\otimes I_{U_1U_2}).
\end{equation}

The balanced beam splitter $B_A$ is specified by
\begin{equation}\label{eq:beam}
 B_A^*(W_{A_+}(s)\otimes W_{A_-}(t))B_A
 =W_{A_1}\!\left(\frac{s+t}{\sqrt2}\right)
  \otimes W_{A_2}\!\left(\frac{s-t}{\sqrt2}\right).
\end{equation}

Let  $S(\eta):=-\Tr(\eta\log\eta)$. Conditional mutual information is
\begin{equation}\label{eq:cmi-definition}
 I(B:C\mid M)_\zeta
 :=S(BM)_\zeta+S(CM)_\zeta-S(M)_\zeta-S(BCM)_\zeta.
\end{equation}
The conditional entropy is
$S(B\mid M)_\zeta:=S(BM)_\zeta-S(M)_\zeta$.

We study
\begin{equation}\label{eq:cmi}
 I(A_+:A_-\mid U_1U_2)_\omega=0.
\end{equation}
Each finite-energy hypothesis gives $S(\rho^{(i)}_{A_i})<\infty$.
Thus all entropies in \eqref{eq:cmi} are finite.

The notation $\simeq$ denotes equality up to a unitary
identification of the relevant Hilbert spaces.
\begin{lemma}[Equality in strong subadditivity]
\label{lem:markov-support}
Let $\zeta_{BMC}$ be a normal state on separable
Hilbert spaces, with
\[
S(B)_\zeta,\ S(M)_\zeta,\ S(C)_\zeta<\infty.
\]
Then $I(B:C\mid M)_\zeta=0$ if and only if there
exists a finite or countable decomposition
\begin{equation}\label{eq:support-markov}
\cH_M\simeq
\bigoplus_a\cH_{L_a}\otimes\cH_{R_a}
\end{equation}
such that, on $\supp\zeta_M$,
\begin{equation}\label{eq:markov-form}
\zeta_{BMC}\simeq
\bigoplus_a r_a\,
\eta^a_{BL_a}\otimes\theta^a_{R_aC},
\end{equation}
where $r_a>0$, $\sum_a r_a=1$, and
$\eta^a_{BL_a}$ and $\theta^a_{R_aC}$
are normal states. The unitary identification
acts only on the conditioning system $M$.
\end{lemma}

\begin{proof}
For faithful joint states, the forward implication
follows from the quantum Markov decomposition of
Jen\v{c}ov\'a and Petz~\cite{JP}.
The extension to arbitrary normal states, together
with the converse implication, is proved in
Appendix~\ref{app:markov}.
\end{proof}

\subsection{Normalized partial characteristic functions}
For each $i\in\{1,2\}$ and $s\in\mathbb R^d$, define the partial characteristic function
\begin{equation}\label{eq:normalization}
 \Phi_i(s):=\Tr_{A_i}[\rho_i(W_{A_i}(s)\otimes I)]
           =\mu_i^{1/2}K_i(s)\mu_i^{1/2},
 \qquad K_i(s)\in\cB(\cH_{U_i}).
\end{equation}

Thus $\Phi_i(s)$ is a trace-class operator on the memory, and
$K_i(s)$ is its bounded normalization on $\supp\mu_i$.
The trace $\Tr_{U_i}\Phi_i(s)$ is the scalar characteristic
function of the physical marginal. The pair $(\mu_i,K_i)$
uniquely determines the joint state $\rho_i$. This normalization allows conditional independence to be expressed through algebraic identities and commutation relations. The following lemma establishes the existence, uniqueness, and boundedness of $K_i(s)$.

\begin{lemma}[Bounded normalization]\label{lem:normalization}
For each $T\in\cB(\cH_A)$ there is a unique $K(T)\in\cB(\supp\mu)$
with $\Tr_A[\rho_{AU}(T\otimes I)]=\mu^{1/2}K(T)\mu^{1/2}$ and
$\norm{K(T)}\le\norm T$. In particular,
\begin{equation}\label{eq:Kbasic}
 K_0=I,\qquad K_{-s}=K_s^*,\qquad\norm{K_s}\le1,
\end{equation}
and $s\mapsto K_s$ is weak-operator continuous.
\end{lemma}

\begin{proof}

We suppress the input index in this single-state argument;
thus $K_s$ denotes $K_i(s)$ for the input under consideration. Set
$\Phi(T):=\Tr_A[\rho_{AU}(T\otimes I)]$.
Cauchy--Schwarz gives
$
|\langle x,\Phi(T)y\rangle|
\le \|T\|\,\|\mu^{1/2}x\|\,\|\mu^{1/2}y\|.
$
Since $\mu$ is faithful, this bound defines a bounded
sesquilinear form $b_T(\mu^{1/2}x,\mu^{1/2}y)
=\langle x,\Phi(T)y\rangle$ on the dense subspace
$\Ran\mu^{1/2}$. By the Riesz representation theorem,
there is a unique $K(T)\in\cB(\cH_U)$ such that
$
\Phi(T)=\mu^{1/2}K(T)\mu^{1/2},
 \|K(T)\|\le\|T\|.
$
Linearity, unitality, and preservation of adjoints
follow by uniqueness. Taking $T=W_A(s)$ yields
$K_0=I$, $K_{-s}=K_s^*$, and $\|K_s\|\le1$. Finally, strong continuity of the Weyl operators implies
that $s\mapsto\Phi(W_A(s))$ is weak-operator continuous.
For vectors $x,y$ in $\Ran\mu^{1/2}$, the matrix elements
of $K_s$ are therefore continuous by the factorization.
These vectors form a dense subspace, and $\|K_s\|\le1$
uniformly. Approximating arbitrary vectors by vectors
in $\Ran\mu^{1/2}$ proves weak-operator continuity
on the full memory space.\end{proof}

We also consider the modular evolution of $K_i(s)$
under the memory state $\mu_i$:
\begin{equation}\label{eq:modular}
 \alpha_u^{(i)}(X):=\mu_i^{\ii u}X\mu_i^{-\ii u},
 \qquad
 K_i^{(u)}(s):=\alpha_u^{(i)}(K_i(s)),
 \qquad X\in\cB(\cH_{U_i}),\quad u\in\R.
\end{equation}
These modular translates will later be used to construct
a real Hilbert space $E$, whose spectral decomposition
under the induced modular dynamics identifies the
bosonic degrees of freedom of the memories.

Consider the following operator identities:
\begin{align}
 K_1(s+t)\otimes K_2(s-t)
 &=K_1(s)K_1(t)\otimes K_2(s)K_2(-t),
 \tag{B}\label{eq:B}\\
 [K_1^{(u)}(s)\otimes K_2^{(u)}(s),
  K_1^{(v)}(t)\otimes K_2^{(v)}(-t)]
 &=0,
 \tag{M}\label{eq:M}
\end{align}
for all $s,t\in\R^d$ and $u,v\in\R$.
We refer to \eqref{eq:B} as the
\emph{operator Bernstein equation} and to
\eqref{eq:M} as the \emph{modular commutation relation}.

\section{Main theorem}

\begin{theorem}[Conditional quantum Kac--Bernstein theorem]
\label{thm:main}
Let $\rho_1$ and $\rho_2$ be two independent normal
states satisfying \eqref{eq:assumptions}.
The following are equivalent:
\begin{enumerate}[label=(\roman*)]

\item $I(A_+:A_-\mid U_1U_2)_\omega=0$.

\item The operator families $K_1$ and $K_2$ satisfy
\eqref{eq:B} and \eqref{eq:M}.

\item There exist a finite or countable set $\mathcal I$,
finite or countable sets $\mathcal J_1,\mathcal J_2$,
and a common centered normal Gaussian state $\gamma_{AQ}$
such that, for $i=1,2$,
\begin{align}
\cH_{U_i}
&\simeq \cH_Q\otimes
\left(\bigoplus_{j\in\mathcal J_i}
\cH_{K_j^{(i)}}\right),
&
\cH_Q&=\cF(\ell^2(\mathcal I;\C)),
\label{eq:Hnormal}\\
\rho_i
&\simeq\bigoplus_{j\in\mathcal J_i}
p_j^{(i)}\gamma_{AQ}^{b_j^{(i)}}\otimes\nu_j^{(i)},
&
\gamma_{AQ}^{b}
&:=(D_A(b)\otimes I)\gamma_{AQ}
(D_A(b)^*\otimes I),
\label{eq:normalform}\\
\gamma_Q
&=\tau_Q:=\bigotimes_{\ell\in\mathcal I}\tau(q_\ell),
&
0<q_\ell&<1,\qquad \sum_\ell q_\ell<\infty.
\label{eq:taunormal}
\end{align}
Here $\simeq$ denotes a unitary identification acting
only on the memory, $p_j^{(i)}>0$ with
$\sum_jp_j^{(i)}=1$, $\nu_j^{(i)}$ is faithful on
$\cH_{K_j^{(i)}}$, and $b_j^{(i)}\in\R^d$.
The set $\mathcal I$ may be empty.
We call $\cH_Q\otimes\cH_{K_j^{(i)}}$
the $j$-th memory sector of input $i$.

\end{enumerate}
\end{theorem}

\begin{proof}[Proof roadmap]
The implication \textup{(i)}$\Rightarrow$\textup{(ii)}
follows from Proposition~\ref{prop:bridge}.
For \textup{(ii)}$\Rightarrow$\textup{(iii)},
Proposition~\ref{prop:classification} classifies the
operator Bernstein equation, Sections~\ref{sec:geometry}
and~\ref{sec:thermal} identify the common bosonic memory
factor, and Subsection~\ref{subsec:gaussian-reconstruction}
establishes the Gaussian normal form.
Finally, Subsection~\ref{subsec:converse} proves
\textup{(iii)}$\Rightarrow$\textup{(i)}
by balanced mixing of the common Gaussian cores.
\end{proof}

Before proceeding to the proof of Theorem~\ref{thm:main},
we present several consequences of the characterization
and an explicit example illustrating the role of
quantum memory and the possibility of entangled
equality states.

\begin{corollary}[Trivial bosonic memory factor]
\label{cor:trivial-Q}
Suppose the common bosonic factor $Q$ in
Theorem~\ref{thm:main} is trivial, equivalently
$\mathcal I=\varnothing$ and $\cH_Q\simeq\C$.
Then the equality states have the form
\begin{equation}\label{eq:finite-normal}
 \rho_i\simeq\bigoplus_{j\in\mathcal J_i}
 p_j^{(i)}\,D_A(b_j^{(i)})\gamma_A
 D_A(b_j^{(i)})^*\otimes\nu_j^{(i)},
 \qquad i=1,2,
\end{equation}
where $\gamma_A$ is a common centered Gaussian state.
In particular, both inputs are separable across $A_i:U_i$,
without any finite-dimensionality assumption on the memories.
\end{corollary}
\begin{proof}
With $\cH_Q\simeq\C$, the common Gaussian core
$\gamma_{AQ}$ in \eqref{eq:normalform} reduces to
$\gamma_A$. The normal form therefore becomes
\eqref{eq:finite-normal}, a convex combination of product
states supported on orthogonal memory subspaces.
\end{proof}

\begin{remark}[Finite-dimensional memory as a special case]
\label{rem:finite-memory}
If $D_i:=\dim\supp\mu_i<\infty$ for at least one
$i\in\{1,2\}$, then the factor $Q$ is necessarily trivial:
a nontrivial bosonic Fock space in \eqref{eq:Hnormal}
is infinite-dimensional. Hence Corollary~\ref{cor:trivial-Q}
applies and both equality inputs have the separable form
\eqref{eq:finite-normal}. Moreover,
$|\mathcal J_i|\le D_i$ for every $i$ with $D_i<\infty$.
An independent direct proof of this finite-dimensional case,
using determinants of the modular commutation relations,
is retained in Subsection~\ref{subsec:finite-memory}.
\end{remark}

\begin{remark}[Common covariance and conditional entropy]
Although $\rho_1$ and $\rho_2$ need not be identical,
the equality condition forces them to share the same
Gaussian core $\gamma_{AQ}$, up to physical displacements.
In particular, they have the same Gaussian covariance,
including the correlations between $A$ and $Q$.

Moreover, the orthogonal memory labels and physical
displacements do not affect conditional entropy. Indeed,
by \eqref{eq:normalform} and the direct-sum entropy formula,
\[
\begin{aligned}
S(A_iU_i)_{\rho_i}
&=H(p^{(i)})+S(\gamma_{AQ})
  +\sum_j p_j^{(i)}S(\nu_j^{(i)}),\\
S(U_i)_{\rho_i}
&=H(p^{(i)})+S(\gamma_Q)
  +\sum_j p_j^{(i)}S(\nu_j^{(i)}).
\end{aligned}\] Subtracting gives
\begin{equation}\label{eq:common-conditional-entropy}
S(A_1\mid U_1)_{\rho_1}
=S(A_2\mid U_2)_{\rho_2}
=S(\gamma_{AQ})-S(\gamma_Q)
=S(A\mid Q)_\gamma.
\end{equation}
Since the entropy of a Gaussian state depends only on
its covariance matrix, the common conditional entropy
is determined entirely by the covariance of
$\gamma_{AQ}$.
\end{remark}

\begin{remark}[Role of the quantum memory]
The decomposition in \eqref{eq:normalform} separates
the memory into two parts. The factor $Q$ carries
the Gaussian correlations with $A$, while the
orthogonal memory sectors distinguish the
different displacements $b_j^{(i)}$. In particular,
any entanglement between $A$ and $U_i$ comes from
the common Gaussian state $\gamma_{AQ}$.
\end{remark}

\begin{example}[Entangled equality states]
\label{ex:entangled-equality}
The equality condition in Theorem~\ref{thm:main}
is compatible with entanglement between the physical
systems and their quantum memories.
To illustrate this, consider the two-mode squeezed
vacuum state
\[
|\psi_r\rangle_{AU}
=\sqrt{1-\lambda^2}
\sum_{m=0}^{\infty}
\lambda^m|m\rangle_A|m\rangle_U,
\qquad
\lambda=\tanh r,\quad r>0.
\]
Take two independent copies,
$\rho_i=|\psi_r\rangle\langle\psi_r|_{A_iU_i}$
for $i=1,2$. Each input is entangled across
$A_i:U_i$, and its memory marginal is the thermal
state $\tau(q)$ with $q=\tanh^2r$.
In our convention, their common Gaussian covariance is
\[
\Gamma_{AU}=
\begin{pmatrix}
\cosh(2r)I_2 & \sinh(2r)Z\\
\sinh(2r)Z & \cosh(2r)I_2
\end{pmatrix},
\qquad Z=\operatorname{diag}(1,-1).
\]

Apply the balanced beam splitter to $A_1A_2$,
and then an identical beam splitter to the memories
$U_1U_2$, producing $U_+$ and $U_-$.
Denote the resulting state by $\widetilde\omega$.
Writing
$\chi_\gamma(x)=\exp(-x^{\mathsf T}\Gamma_{AU}x/4)$
for the characteristic function of the common
Gaussian input, we obtain
\[
\begin{aligned}
\chi_{\widetilde\omega}(x,y)
&=\chi_\gamma\left(\frac{x+y}{\sqrt2}\right)
  \chi_\gamma\left(\frac{x-y}{\sqrt2}\right)
=\chi_\gamma(x)\chi_\gamma(y),
\end{aligned}
\]
where $x$ and $y$ denote the phase-space coordinates
of $A_+U_+$ and $A_-U_-$, respectively.
The second equality follows from the parallelogram
identity for quadratic forms.
By uniqueness of Weyl characteristic functions,
\[
\widetilde\omega
=
|\psi_r\rangle\langle\psi_r|_{A_+U_+}
\otimes
|\psi_r\rangle\langle\psi_r|_{A_-U_-}.
\]
Thus $I(A_+:A_-\mid U_+U_-)_{\widetilde\omega}=0$.
Since the additional beam splitter acts only on
the conditioning memory, invariance of conditional
mutual information under memory-local unitaries gives
\[
I(A_+:A_-\mid U_1U_2)_\omega=0.
\]
Hence the two inputs provide entangled equality states
with a nontrivial common bosonic memory factor.
\end{example}

\section{The operator Bernstein equation}
\label{sec:bernstein}

We begin by deriving the operator Bernstein equation
\eqref{eq:B} and the modular commutation relation
\eqref{eq:M} from the conditional-independence condition.
The key step is to combine the Markov decomposition of
the output state with the action of the beam splitter
on Weyl operators.

\begin{proposition}[Markov reduction]\label{prop:bridge}
Under the assumptions of Theorem~\ref{thm:main},
condition \textup{(i)} implies \eqref{eq:B} and \eqref{eq:M}.
\end{proposition}

\begin{proof}
By Lemma~\ref{lem:markov-support},
$I(A_+:A_-\mid U_1U_2)_\omega=0$ gives
\begin{equation}\label{eq:markov-structure}
\cH_{U_1U_2}\simeq\bigoplus_a\cH_{L_a}\otimes\cH_{R_a},
\qquad
\omega\simeq\bigoplus_a r_a\,
\eta^a_{A_+L_a}\otimes\theta^a_{R_aA_-}.
\end{equation}
We work in this representation, with the unitary
identification acting only on the memory.

Let $F(s,t)$ be the partial characteristic function
of $\omega$, normalized with respect to
$\mu_1\otimes\mu_2$, as
\[
\begin{aligned}
&\Tr_{A_+A_-}
[\omega(W_{A_+}(s)\otimes W_{A_-}(t)\otimes I)]
=(\mu_1\otimes\mu_2)^{1/2}
F(s,t)(\mu_1\otimes\mu_2)^{1/2}.
\end{aligned}
\]
Set $F_+(s):=F(s,0)$ and $F_-(t):=F(0,t)$.

To apply the normalization blockwise, note that the memory
marginal in \eqref{eq:markov-structure} is
$\mu_1\otimes\mu_2\simeq\bigoplus_a r_a\mu_{L_a}^a\otimes\mu_{R_a}^a$,
where $\mu_{L_a}^a$ and $\mu_{R_a}^a$ are the memory marginals
of $\eta^a$ and $\theta^a$. Write $K_{L_a}(s)$
and $K_{R_a}(t)$ for the bounded normalizations
of the partial Weyl coefficients of $\eta^a$ and
$\theta^a$ relative to $\mu_{L_a}^a$ and
$\mu_{R_a}^a$, respectively, as in
Lemma~\ref{lem:normalization}.
In the $a$th block, the normalized coefficient
of the $A_+$ Weyl operator acts as
$K_{L_a}(s)\otimes I$, while the coefficient of the
$A_-$ operator acts as $I\otimes K_{R_a}(t)$.
The two coefficients therefore multiply to the
normalized joint coefficient in that block, with no
additional factor of $r_a$. Taking their direct sum gives

\begin{equation}\label{eq:markov-coefficients}
F(s,t)=F_+(s)F_-(t),\qquad
[\alpha_u(F_+(s)),\alpha_v(F_-(t))]=0.
\end{equation}

Here $\alpha_u:=\alpha_u^{(1)}\otimes\alpha_u^{(2)}$ is the modular automorphism defined by \eqref{eq:modular}. Within each block, the modular action is induced by
$\mu_{L_a}^a\otimes\mu_{R_a}^a$: it maps
$\cB(\cH_{L_a})\otimes I$ and
$I\otimes\cB(\cH_{R_a})$ into themselves.
Consequently the modular translates of the two
normalized endpoint coefficients commute, even when
the modular parameters $u$ and $v$ differ.

The beam-splitter relation \eqref{eq:beam} gives

\begin{equation}\label{eq:output-coefficients}
\begin{aligned}
F(s,t)&=K_1((s+t)/\sqrt2)\otimes K_2((s-t)/\sqrt2),\\
F_+(s)&=K_1(s/\sqrt2)\otimes K_2(s/\sqrt2),
\qquad
F_-(t)=K_1(t/\sqrt2)\otimes K_2(-t/\sqrt2).
\end{aligned}
\end{equation}

Substitution into the product identity in
\eqref{eq:markov-coefficients}, followed by
$(s,t)\mapsto(\sqrt2\,s,\sqrt2\,t)$, gives
\eqref{eq:B}. Substitution into the commutator identity,
with the same rescaling, gives \eqref{eq:M}.

\end{proof}

\begin{remark}
The normalization in \eqref{eq:normalization} is
essential here. It removes the memory marginal from
the partial characteristic functions, so that
$F_+(s)$ and $F_-(t)$ act on the respective tensor
factors $L_a$ and $R_a$ in each Markov block.
Without this normalization, the memory density
factors would remain, and the product relation
in \eqref{eq:markov-coefficients} would not hold
in this form.
\end{remark}

\subsection{Classification of the operator Bernstein equation}

We now characterize the operator families satisfying
\eqref{eq:B}. As in the classical Bernstein theorem,
the equation forces a common Gaussian factor. In the
quantum setting, the remaining factors are unitary
representations that may satisfy nontrivial commutation
relations. The additional condition \eqref{eq:M}
also constrains their modular translates.

\begin{proposition}[Paired Bernstein classification]
\label{prop:classification}
Let $K_i:\R^d\to\cB(\cH_{U_i})$ be weakly continuous
families satisfying \eqref{eq:Kbasic} and \eqref{eq:B}.
Then there exist unique matrices
$G=G^{\mathsf T}\succeq0$ and
$\Theta=-\Theta^{\mathsf T}$, and strongly continuous
unitary families $V_i(s)$ such that, for $i=1,2$,
\begin{equation}\label{eq:classification}
K_i(s)=e^{-s^{\mathsf T}Gs/4}V_i(s),\qquad
V_i(s)V_i(t)
=e^{-\ii s^{\mathsf T}\Theta t/2}V_i(s+t).
\end{equation}
If \eqref{eq:M} also holds, there are real matrices
$C_{uv}$, common to both families, such that
\begin{equation}\label{eq:translated-ccr}
V_i^{(u)}(s)V_i^{(v)}(t)
=e^{-\ii s^{\mathsf T}C_{uv}t}
V_i^{(v)}(t)V_i^{(u)}(s),
\end{equation}
where $V_i^{(u)}(s):=\alpha_u^{(i)}(V_i(s))$.
Moreover,
\[
C_{uu}=\Theta,\qquad
C_{vu}=-C_{uv}^{\mathsf T},\qquad
C_{u+w,v+w}=C_{uv}.
\]
\end{proposition}

\begin{proof}

We first show that both families have the same strictly
positive scalar modulus. We then identify the scalar
multipliers of their unitary parts and finally apply
the modular commutation relation.

We use the following fact, which we
refer to as \emph{tensor proportionality}:
if $A\otimes B=C\otimes D\ne0$, where
$A,C\in\cB(\cH_1)$ and $B,D\in\cB(\cH_2)$,
then $A=rC$ and $B=r^{-1}D$ for some
$r\in\C\setminus\{0\}$.
\phantomsection\label{fact:tensor-proportionality}

Setting $s=t=h$ and $s=-t=h$ in \eqref{eq:B}
gives
\begin{equation}\label{eq:paired-doubling}
\begin{aligned}
K_1(2h)\otimes I
&=K_1(h)^2\otimes K_2(h)K_2(h)^*, \quad
I\otimes K_2(2h)
=K_1(h)K_1(h)^*\otimes K_2(h)^2.
\end{aligned}
\end{equation}
Weak continuity at zero ensures that both left-hand
sides are nonzero for sufficiently small $h$.
Applying tensor proportionality, also with $-h$,
shows that
\[
K_i(h)K_i(h)^*=K_i(h)^*K_i(h)=r_i(h)^2I,
\qquad r_i(h)>0.
\]
Thus $K_i(h)=r_i(h)V_i(h)$ with $V_i(h)$
unitary. Taking norms in \eqref{eq:paired-doubling}
yields
\[
r_1(2h)=r_2(2h)=r_1(h)^2r_2(h)^2.
\]

For any $s\in\R^d$, choose an integer $k$ so large
that $h=2^{-k}s$ lies in the neighborhood where the
preceding argument applies. The doubling identities
express $K_i(2h)$ as a nonzero scalar multiple of a
product of unitaries and hence extend the decomposition
successively to $2h,4h,\ldots,s$. At each doubling,
the norms of $K_1(2h)$ and $K_2(2h)$ agree.
Applying this equality at $h=s/2$, after extension,
gives $r_1(s)=r_2(s)=:r(s)>0$.

Since $r(s)=\|K_i(s)\|\le1$, taking norms in
\eqref{eq:B} gives
\[
q(s+t)+q(s-t)=2q(s)+2q(t),
\qquad q(s):=-\log r(s)\ge0.
\]
By the classical quadratic functional equation
\cite[Section~4.1]{Kannappan},
$q(s)=s^{\mathsf T}Gs/4$ for a unique real
symmetric matrix $G\succeq0$. Consequently $
K_i(s)=e^{-s^{\mathsf T}Gs/4}V_i(s),
$ where the unitary families $V_i$ are strongly
continuous.

Since the scalar Gaussian factors obey the quadratic
identity for $q$, they cancel from \eqref{eq:B}.
Tensor proportionality can then be applied to the
unitary factors, yielding

\begin{equation}\label{eq:projective-multiplier}
V_i(s)V_i(t)=c_i(s,t)V_i(s+t),
\qquad c_1(s,t)c_2(s,-t)=1,
\end{equation}
where $c_i$ are continuous scalar multipliers.

We next show that the multipliers are antisymmetric
in their arguments. Taking adjoints in the product
identity and using $V_i(-s)=V_i(s)^*$ gives
$c_i(-t,-s)=\overline{c_i(s,t)}$.
For the second family, replacing $t$ by $-t$ in this
identity gives
$c_2(t,-s)=\overline{c_2(s,-t)}$.
Applying the paired relation in
\eqref{eq:projective-multiplier} first to $(s,t)$
and then to $(t,s)$ therefore gives
\[
 c_1(t,s)=c_2(t,-s)^{-1}
 =c_2(s,-t)=c_1(s,t)^{-1}.
\]
The same argument with the two families interchanged
(and the sign of $t$ reversed) yields
$c_2(t,s)=c_2(s,t)^{-1}$.

The scalar commutator of each unitary family is
$c_i(s,t)/c_i(t,s)=c_i(s,t)^2$.
The cocycle identity implies that this commutator
is multiplicative in each argument:
\[
c_i(s+r,t)^2=c_i(s,t)^2c_i(r,t)^2,\qquad
c_i(s,t+r)^2=c_i(s,t)^2c_i(s,r)^2.
\]
Hence, as a continuous bicharacter of $\R^d$,
$c_i(s,t)^2=e^{-\ii s^{\mathsf T}\Theta_i t}$
for some real antisymmetric matrix $\Theta_i$.
Since
$\left(c_i(s,t)e^{\ii s^{\mathsf T}\Theta_i t/2}\right)^2=1$,
continuity, connectedness of $\R^{2d}$, and
$c_i(0,t)=1$ imply
\[
c_i(s,t)=e^{-\ii s^{\mathsf T}\Theta_i t/2}.
\]

Equation~\eqref{eq:projective-multiplier} then gives
$\Theta_1=\Theta_2=:\Theta$, proving
\eqref{eq:classification}. Substituting this
classification into \eqref{eq:M} and canceling
the scalar factors yields
\[
[V_1^{(u)}(s)\otimes V_2^{(u)}(s),
 V_1^{(v)}(t)\otimes V_2^{(v)}(-t)]=0.
\]

Tensor proportionality gives inverse commutation
phases for the two tensor factors in \eqref{eq:M}.
Since the second factor contains $V_2^{(v)}(-t)
=V_2^{(v)}(t)^*$, both families have the same
commutation phase when expressed at $(s,t)$. To verify multiplicativity, write this common phase as
$\chi_{uv}(s,t)$, so that
$V_i^{(u)}(s)V_i^{(v)}(t)
=\chi_{uv}(s,t)V_i^{(v)}(t)V_i^{(u)}(s)$.
The modularly translated families retain the projective
product law \eqref{eq:classification}, with the same scalar
multiplier. Commuting $V_i^{(v)}(t)$ past the product
$V_i^{(u)}(s)V_i^{(u)}(r)$, and canceling that multiplier,
gives the first identity below. Commuting $V_i^{(u)}(s)$
past $V_i^{(v)}(t)V_i^{(v)}(r)$ gives the second:
\[
 \begin{aligned}
 \chi_{uv}(s+r,t)&=\chi_{uv}(s,t)\chi_{uv}(r,t), \quad
 \chi_{uv}(s,t+r)=\chi_{uv}(s,t)\chi_{uv}(s,r).
 \end{aligned}
\]
Thus the continuous phase is a bicharacter of $\R^d$,
and has the form
$\chi_{uv}(s,t)=e^{-\ii s^{\mathsf T}C_{uv}t}$
for a real matrix $C_{uv}$. Setting $u=v$, interchanging the modular parameters, and applying a common modular
translation yield, respectively,\[
C_{uu}=\Theta,\qquad
C_{vu}=-C_{uv}^{\mathsf T},\qquad
C_{u+w,v+w}=C_{uv}.\]
\end{proof}
\begin{remark}

The matrix $G$ in \eqref{eq:classification} is the
covariance of the Gaussian part of the physical
system that is independent of the memory. It is
common to both inputs, but is not the full physical
covariance of the joint Gaussian core. The latter
also includes the contribution from the bosonic
memory, which will be determined by the modular
analysis.

\end{remark}

\subsection{The finite-dimensional memory case}
\label{subsec:finite-memory}

We retain a direct proof of the finite-dimensional
special case in Remark~\ref{rem:finite-memory}, independent
of the Fock-factor argument used for
Corollary~\ref{cor:trivial-Q}. Suppose, for example,
that $\dim\cH_{U_1}=m<\infty$.

The modular commutation relations give
\[
V_1^{(u)}(s)V_1^{(v)}(t)
=e^{-\ii s^{\mathsf T}C_{uv}t}
V_1^{(v)}(t)V_1^{(u)}(s).
\]
Taking determinants yields
$e^{-\ii m s^{\mathsf T}C_{uv}t}=1$
for all $s,t\in\R^d$ and $u,v\in\R$.
Continuity in $s,t$ implies $C_{uv}=0$
for every $u,v$. Since $C_{uv}$ is common
to both inputs, all modular translates of
$V_i$ commute for $i=1,2$.
In particular, taking $u=v=0$ gives
$C_{00}=\Theta=0$, so each $V_i$ is
a commuting unitary representation.

Let $\mathcal A_i$ be the von Neumann algebra generated
by $\{\alpha_u^{(i)}(V_i(s)):u\in\R,\ s\in\R^d\}$.
All its generators commute, so $\mathcal A_i$ is abelian;
it is modularly invariant because applying
$\alpha_v^{(i)}$ shifts $u$ to $u+v$.
The restriction of the faithful normal state $\mu_i$
to $\mathcal A_i$ is faithful and normal. By the
modular-invariance theorem \cite{Takesaki}, the restriction of
$\alpha_u^{(i)}$ to $\mathcal A_i$ is the modular
group of this restricted state. A faithful normal
state on an abelian von Neumann algebra has trivial
modular group. Therefore $\alpha_u^{(i)}$ fixes
every element of $\mathcal A_i$, and in particular
$\mu_i^{\ii u}V_i(s)\mu_i^{-\ii u}=V_i(s)$
for all $s,u$. Hence $[V_i(s),\mu_i]=0$.

Because $\mu_i$ is faithful and trace class,
its eigenspaces are finite-dimensional.
The commuting operators $V_i(s)$ preserve
these eigenspaces and can be simultaneously
diagonalized on each of them. Grouping equal
characters gives
\[
\cH_{U_i}
=\bigoplus_{j\in\mathcal J_i}\cH_{K_j^{(i)}},
\qquad
V_i(s)=\bigoplus_j
e^{\ii(b_j^{(i)})^{\mathsf T}s}I,
\qquad
\mu_i=\bigoplus_jp_j^{(i)}\nu_j^{(i)}.
\]

Since $K_i(s)=e^{-s^{\mathsf T}Gs/4}V_i(s)$,
the partial characteristic function is
\[
\Phi_i(s)=\bigoplus_j p_j^{(i)}
e^{-s^{\mathsf T}Gs/4+
\ii(b_j^{(i)})^{\mathsf T}s}\nu_j^{(i)}.
\]
Taking the trace over the memory gives the
physical characteristic function
\[
\chi_{\rho_{A_i}}(s)
=\sum_j p_j^{(i)}
e^{-s^{\mathsf T}Gs/4+
\ii(b_j^{(i)})^{\mathsf T}s}.
\]
Weyl uniqueness applied to the partial
coefficients therefore gives
\[
\rho_i\simeq\bigoplus_j p_j^{(i)}
\gamma_A^{b_j^{(i)}}\otimes\nu_j^{(i)},
\]
where $\gamma_A$ is the centered Gaussian
state with covariance $G$, common to both inputs.

It remains to treat the case in which both
memories are infinite-dimensional. We now
develop the modular geometry needed to
identify their common Gaussian memory factor.

\section{Modular geometry and the common oscillator structure}
\label{sec:geometry}

Throughout this section, we fix either input and suppress
the index $i$, writing $\rho,\mu,K,V,\alpha_u$.
The modular translates of $V(s)$ generate memory
observables whose commutation relations are determined
by \eqref{eq:translated-ccr}. We use these relations
to construct a real covariance space carrying a Weyl
representation and an induced modular action. The
spectral decomposition of this action will identify
the bosonic degrees of freedom of the memories.

\subsection{The covariance space of the memory fields}
\label{subsec:fields}

For $k=1,\ldots,d$, the Weyl relations
\eqref{eq:classification} imply that
$t\mapsto V(te_k)$ is a strongly continuous
one-parameter unitary group. By Stone's theorem
\cite[Section~3.3]{Derezinski}, there is a
self-adjoint generator $A_k$ such that
$V(te_k)=e^{\ii tA_k}$.

Define
\begin{equation}\label{eq:centered-fields}
b_0:=\Tr(\rho_AR_A),\qquad
B_{k,u}:=\mu^{\ii u}(A_k-b_{0,k}I)\mu^{-\ii u}.
\end{equation}
The finite-energy assumption gives
\begin{equation}\label{eq:moment-transfer}
\Tr(\mu A_k)=b_{0,k},\qquad
\Tr(\mu A_k^2)
=\Tr(\rho_AR_{A,k}^2)-\tfrac12G_{kk}<\infty.
\end{equation}
The moment identities are justified in
Appendix~\ref{app:fields:moments};
traces of unbounded operators are understood
as spectral moments.

Let $\mathcal D$ be the free real vector space
generated by symbols $\varepsilon_{k,u}$,
$1\le k\le d$, $u\in\R$, corresponding to
the fields $B_{k,u}$. Every $a\in\mathcal D$
has a finite expansion
$a=\sum_{m=1}^{M}a_m\varepsilon_{k_m,u_m}$. 
The commutation relations
\eqref{eq:translated-ccr} determine an
antisymmetric bilinear form $\sigma$ on
$\mathcal D$ by
\begin{equation}\label{eq:modular-commutation-form}
\sigma(\varepsilon_{k,u},\varepsilon_{l,v})
:=(C_{uv})_{kl}.
\end{equation}

For $a\in\mathcal D$, let $\mathscr W(a)$
denote the phase-corrected product of the
unitaries $e^{\ii a_mB_{k_m,u_m}}$.
As detailed in Appendix~\ref{app:fields:domains},
these operators satisfy the Weyl relations
$\mathscr W(a)\mathscr W(b)
=e^{-\ii\sigma(a,b)/2}\mathscr W(a+b)$.
By Stone's theorem, the strongly continuous
group $t\mapsto\mathscr W(ta)$ has a
self-adjoint generator $F_a$, satisfying
$\mathscr W(ta)=e^{\ii tF_a}$.

Define the covariance form
\begin{equation}\label{eq:covariance-form}
\Lambda(a):=F_a\mu^{1/2},\qquad
g(a,b):=\Rea
\ip{\Lambda(a)}{\Lambda(b)}_{\HS},
\end{equation}
where $\ip{X}{Y}_{\HS}:=\Tr(X^*Y)$.
The finite second moments imply
$\Lambda(a)\in\HS$ for every $a\in\mathcal D$;
see Appendix~\ref{app:fields:domains}.
The following lemma establishes the corresponding
completed Weyl and modular structures.

\begin{lemma}[Modular covariance space]
\label{lem:fields}
The completion $E:=\overline{\mathcal D/\ker g}$
is a separable real Hilbert space, and $\sigma$
induces a continuous antisymmetric form $\sigma_E$
on $E$. There exist a strongly continuous Weyl family
$\mathscr W:E\to\mathcal U(\cH_U)$, a strongly
continuous orthogonal group $T(u)$ on $E$, and
a real-linear map $\iota:\R^d\to E$ such that
\begin{align}
\mathscr W(a)\mathscr W(b)
&=e^{-\ii\sigma_E(a,b)/2}\mathscr W(a+b),
&
V(s)&=e^{\ii b_0^{\mathsf T}s}\mathscr W(\iota s),
\label{eq:field-weyl}\\
\alpha_u(\mathscr W(a))
&=\mathscr W(T(u)a),
&
E&=\overline{\operatorname{span}_{\R}
\{T(u)\iota s:u\in\R,\ s\in\R^d\}}.
\label{eq:field-orbits}
\end{align}
For $\mathscr W(ta)=e^{\ii tF_a}$, we have
$\Ran\mu^{1/2}\subset\Dom F_a$ and
$\Lambda(a)=F_a\mu^{1/2}\in\HS$, with
\begin{equation}\label{eq:field-covariance}
\Tr(\mu F_a)=0,\qquad
\ip{\Lambda(a)}{\Lambda(b)}_{\HS}
=g(a,b)+\tfrac{\ii}{2}\sigma_E(a,b).
\end{equation}
\end{lemma}

The construction relies on the estimates
\begin{equation}\label{eq:field-estimates}
|\sigma_E(a,b)|\le2\norm a_E\norm b_E,\qquad
\norm{(\mathscr W(a)-I)\mu^{1/2}}_{\HS}
\le\norm a_E.
\end{equation}
The first follows from the covariance--commutator
identity, while the second gives the strong extension
of the Weyl representation using faithfulness of $\mu$.
The domain and completion arguments are provided in
Appendices~\ref{app:fields:covariance} and
\ref{app:fields:completion}.
The modular shift
$\varepsilon_{k,u}\mapsto\varepsilon_{k,u+t}$
induces $T(t)$, with
$\iota e_k=[\varepsilon_{k,0}]$.

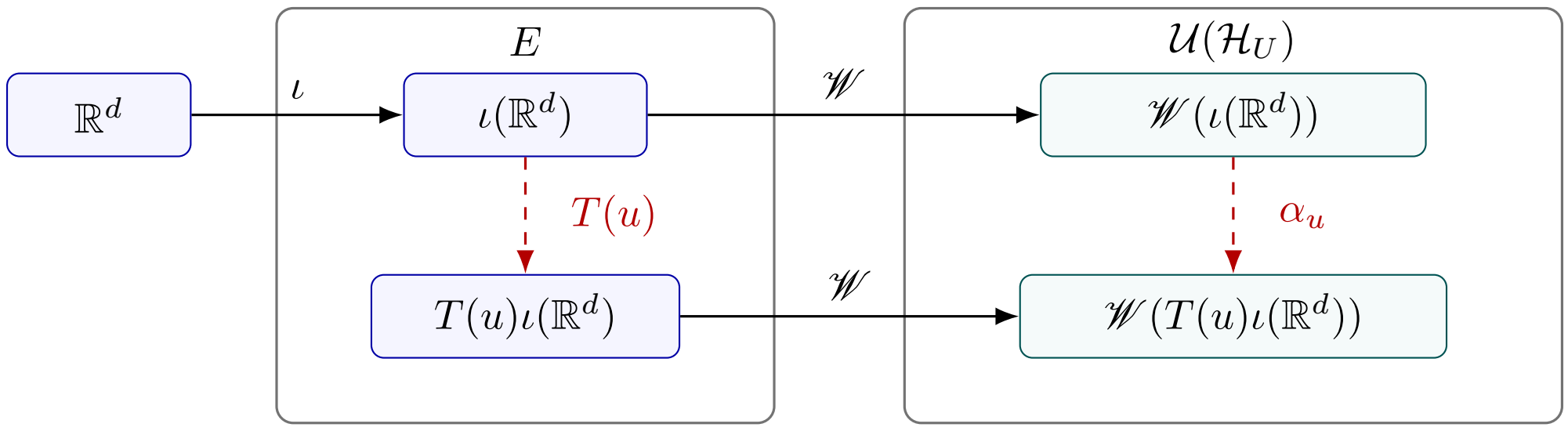
\begin{figure}[htbp]
\centering
\begin{tikzpicture}[
 font=\small, >=Latex,
 mainarrow/.style={-{Latex[length=2mm]},semithick},
 modular/.style={-{Latex[length=2mm]},semithick,dashed,red!70!black},
 ambient/.style={draw=black!55,rounded corners=4pt,semithick},
 seed/.style={draw=blue!65!black,fill=blue!4,
 rounded corners=3pt,minimum width=2.05cm,
 minimum height=.7cm,align=center},
 image/.style={draw=teal!65!black,fill=teal!4,
 rounded corners=3pt,minimum width=3.25cm,
 minimum height=.7cm,align=center}
]
\draw[ambient] (-3.65,-1.75) rectangle (0.55,1.75);
\draw[ambient] (1.65,-1.75) rectangle (7.2,1.75);

\node at (-1.55,1.47) {$E$};
\node at (4.425,1.47) {$\mathcal U(\cH_U)$};

\node[seed,minimum width=1.55cm]
(R) at (-5.15,.85) {$\R^d$};
\node[seed] (I) at (-1.55,.85)
{$\iota(\R^d)$};
\node[seed,minimum width=2.6cm]
(TI) at (-1.55,-.85)
{$T(u)\iota(\R^d)$};

\node[image] (WI) at (4.425,.85)
{$\mathscr W(\iota(\R^d))$};
\node[image,minimum width=3.6cm]
(WTI) at (4.425,-.85)
{$\mathscr W(T(u)\iota(\R^d))$};

\draw[mainarrow] (R.east)
--node[above] {$\iota$} (I.west);
\draw[mainarrow] (I.east)
--node[above] {$\mathscr W$} (WI.west);
\draw[mainarrow] (TI.east)
--node[above] {$\mathscr W$} (WTI.west);

\draw[modular] (I.south)
--node[right=7pt] {$T(u)$} (TI.north);
\draw[modular] (WI.south)
--node[right=7pt] {$\alpha_u$} (WTI.north);
\end{tikzpicture}

\caption{Modular covariance-space construction.
The initial directions $\iota(\R^d)$ and
their modular translates $T(u)\iota(\R^d)$
generate $E$ by closed real linear span.
The Weyl representation $\mathscr W$
intertwines the orthogonal transformations
$T(u)$ with the memory modular automorphisms
$\alpha_u$.}
\label{fig:modular-covariance}
\end{figure}

Figure~\ref{fig:modular-covariance} illustrates
the Weyl representation and its compatibility
with the modular action.


\begin{remark}[CCR representation and modular dynamics]

The relations \eqref{eq:field-weyl} are the
standard Weyl form of the canonical commutation
relations; see \cite[Theorem~2.2]{PetzCCR},
whose symplectic convention corresponds to
$\sigma_{\mathrm P}=-\sigma_E/2$.
The cited result establishes the associated
abstract CCR algebra, while
Lemma~\ref{lem:fields} gives its strongly
continuous realization on $\cH_U$ with the
required covariance and domain properties.
The modular automorphisms $\alpha_u$ induce
the orthogonal dynamics $T(u)$ on $E$.

\end{remark}

\subsection{Fixed and rotating memory fields}
\label{subsec:balance}

We analyze the modular action $T(u)$ on $E$.
Since $\mu$ is faithful and trace class, its
modular group has pure-point spectrum on
$\HS(\cH_U)$. This induces a frequency
decomposition of $E$ and determines the relation
between its covariance and commutation forms.

\begin{lemma}[Spectral balance]
\label{lem:balance}
There exist a countable set
$\mathcal W\subset(0,\infty)$ and a real
orthogonal decomposition
\begin{equation}\label{eq:frequency-decomposition}
E=E_0\oplus\bigoplus_{\omega\in\mathcal W}E_\omega,
\qquad
T(u)|_{E_\omega}
=\cos(\omega u)I+\sin(\omega u)J_\omega,
\end{equation}
where $J_\omega$ is an orthogonal complex
structure on $E_\omega$, with $J_\omega^2=-I$.
Moreover,
$E_0=\rad\sigma_E
:=\{a\in E:\sigma_E(a,b)=0
\text{ for all }b\in E\}$.
Distinct frequency subspaces are
$\sigma_E$-orthogonal, and
\begin{equation}\label{eq:balance}
\sigma_E(a,b)=\kappa_\omega g(a,J_\omega b),
\qquad
\kappa_\omega:=2\tanh(\omega/2),
\quad a,b\in E_\omega.
\end{equation}
Furthermore, $\dim E_0\le d$ and
$\dim E_\omega\le2d$.
\end{lemma}

\begin{proof}
Choose an orthonormal eigenbasis $\{e_r\}$
of $\mu$, with $\mu e_r=u_re_r$ and $u_r>0$.
The modular action on matrix units is
\begin{equation}\label{eq:matrix-modular}
\alpha_u(|e_r\rangle\langle e_s|)
=e^{\ii u\omega_{rs}}|e_r\rangle\langle e_s|,
\qquad
\omega_{rs}:=\log u_r-\log u_s.
\end{equation}
Since $\Lambda:E\to\HS(\cH_U)$ is an
isometry intertwining $T(u)$ and $\alpha_u$,
the pure-point decomposition of the latter
induces the corresponding decomposition of $E$.

The mean-ergodic theorem
\cite[Chapter~2]{Krengel} gives
\begin{equation}\label{eq:frequency-projections}
\begin{aligned}
P_0a
&=\lim_{T\to\infty}\frac1{2T}
\int_{-T}^T T(u)a\,du,\\
P_\omega a
&=\lim_{T\to\infty}\frac1T
\int_{-T}^T\cos(\omega u)T(u)a\,du,
\qquad\omega>0,
\end{aligned}
\end{equation}
where the limits are taken in $E$.
By \eqref{eq:matrix-modular}, these projections
select the zero-frequency and paired
$\pm\omega$ components, respectively.
Their ranges give
\eqref{eq:frequency-decomposition}, with
$E_0=P_0E$, $E_\omega=P_\omega E$, and
$J_\omega=T(\pi/(2\omega))|_{E_\omega}$.
In particular, $J_\omega^2=-I$.

We next establish \eqref{eq:balance}.
The relation is the frequency-wise form of
the standard KMS covariance--commutator
identity (see, e.g., \cite{BratteliRobinson});
we verify its normalization directly.

For $a,b\in E_\omega$, write
$A=F_a$, $B=F_b$, and
$A_{rs}:=\langle e_r,Ae_s\rangle$,
$B_{rs}:=\langle e_r,Be_s\rangle$.
By \eqref{eq:covariance-form},
\[
\ip{\Lambda(a)}{\Lambda(b)}_{\HS}
=\sum_{r,s}u_s\overline{A_{rs}}B_{rs}.
\]
For each pair $(r,s),(s,r)$ satisfying
$u_r=e^\omega u_s$, put
$z_{rs}:=\overline{A_{rs}}B_{rs}$.
The corresponding terms are
$u_sz_{rs}+u_r\overline{z_{rs}}$, giving
\[
\begin{aligned}
g(a,b)\big|_{r,s}
&=(u_r+u_s)\Rea z_{rs},\\
\frac{\sigma_E(a,b)}2\bigg|_{r,s}
&=(u_s-u_r)\Ima z_{rs}.
\end{aligned}
\]

By modular covariance,
$\Lambda(J_\omega b)
=\alpha_{\pi/(2\omega)}(\Lambda(b))$.
The frequencies $\pm\omega$ acquire
phases $\pm\ii$, and hence the paired
contribution to $g(a,J_\omega b)$ is
$-(u_r+u_s)\Ima z_{rs}$.
Since
$(u_r-u_s)/(u_r+u_s)=\tanh(\omega/2)$,
summing over the matrix entries yields
\eqref{eq:balance}.

For zero-frequency entries, $u_r=u_s$,
so $\sigma_E$ vanishes on $E_0$.
Distinct frequency subspaces are
$\sigma_E$-orthogonal.
Moreover, \eqref{eq:balance} gives
\begin{equation}\label{eq:balance-norm}
\sigma_E(a,J_\omega a)
=-\kappa_\omega\norm a_E^2,
\qquad a\in E_\omega.
\end{equation}
Thus $\sigma_E$ is nondegenerate on
each $E_\omega$, $\omega>0$, and
$E_0=\rad\sigma_E$.

Finally, by \eqref{eq:field-orbits},
$E$ is generated by the modular translates
of $\iota e_k$, $1\le k\le d$.
Since $P_\omega$ commutes with $T(u)$,
each $E_\omega$ is spanned by
$P_\omega\iota e_k$ and
$J_\omega P_\omega\iota e_k$.
Likewise, $E_0$ is spanned by
$P_0\iota e_k$.
Consequently, $\dim E_0\le d$ and
$\dim E_\omega\le2d$.
\end{proof}

By \eqref{eq:balance-norm}, each
positive-frequency subspace is symplectic
and therefore admits canonical coordinates.
The corresponding modular frequencies
determine the thermal parameters
$q_\ell=e^{-\omega_\ell}$ used in the
following section.

\subsection{Identification of the rotating spaces}

We now restore the input index $i=1,2$.
Although the covariance forms $g_i$ may differ,
the two inputs share the commutation matrices $C_{uv}$
by Proposition~\ref{prop:classification}.
Together with \eqref{eq:balance}, this determines
a common geometry on their rotating subspaces.

\begin{proposition}[Common rotating geometry]
\label{prop:alignment}
Let $E_i^{\mathrm{rot}}:=E_{i,0}^{\perp}$ and
$P_{i,\mathrm{rot}}:=I-P_{i,0}$.
There exists an onto real-linear isometry
$\mathcal S:E_1^{\mathrm{rot}}\to E_2^{\mathrm{rot}}$
such that
\begin{equation}\label{eq:alignment}
\begin{aligned}
\sigma_{E_2}(\mathcal Sa,\mathcal Sb)
&=\sigma_{E_1}(a,b),&
\mathcal S T_1(u)&=T_2(u)\mathcal S,\\
\mathcal S P_{1,\mathrm{rot}}\iota_1s
&=P_{2,\mathrm{rot}}\iota_2s,
\end{aligned}
\end{equation}
for $a,b\in E_1^{\mathrm{rot}}$,
$s\in\R^d$, and $u\in\R$.
In particular, both inputs have the same nonzero
modular frequencies, with equal multiplicities.
\end{proposition}

\begin{proof}
Let $\pi_i:\mathcal D\to E_i$ be the canonical maps.
Since both commutation forms are induced by the
same $\sigma$ on $\mathcal D$,
\[
\sigma_{E_i}(\pi_i a,\pi_i b)=\sigma(a,b),
\qquad i=1,2.
\]
The modular actions are also induced by the same
shift on $\mathcal D$. Thus, applying the spectral
averages \eqref{eq:frequency-projections} gives
\[
\begin{aligned}
&\sigma_{E_1}(P_{1,\omega}\pi_1a,
J_{1,\omega}P_{1,\omega}\pi_1b)\\
&\qquad=
\sigma_{E_2}(P_{2,\omega}\pi_2a,
J_{2,\omega}P_{2,\omega}\pi_2b).
\end{aligned}
\]

For $\omega>0$, \eqref{eq:balance-norm}
expresses the inner product $g_i$ on
$E_{i,\omega}$ by the commutation form paired
with $J_{i,\omega}$, with the same factor for
both inputs. Hence the preceding equality first
identifies the squared norms of the projected
vectors. Applying real polarization to these
norm equalities yields

\[
g_1(P_{1,\omega}\pi_1a,P_{1,\omega}\pi_1b)
=
g_2(P_{2,\omega}\pi_2a,P_{2,\omega}\pi_2b).
\]

In particular, a zero-norm vector on the first
side has a zero-norm counterpart on the second,
so the assignment
$P_{1,\omega}\pi_1a\mapsto P_{2,\omega}\pi_2a$
is well-defined, including on linear combinations.
Its image is dense because the projected images of
$\mathcal D$ span a dense subspace of $E_{2,\omega}$.
Thus it extends to an onto isometry

$\mathcal S_\omega:E_{1,\omega}\to E_{2,\omega}$.
It preserves the commutation forms and intertwines
the modular actions by construction. Taking
$\mathcal S:=\bigoplus_{\omega>0}\mathcal S_\omega$
proves \eqref{eq:alignment}.
\end{proof}

\section{Thermal realization inside the memories}
\label{sec:thermal}

Proposition~\ref{prop:alignment} identifies the rotating
spaces of both inputs, including their modular frequencies
and commutation forms. We now use this common structure
to construct bosonic modes inside each memory.
For finitely many modes, the required tensor factorization
follows from the representation theory of the canonical
commutation relations. For countably many modes, we
must additionally show that the resulting Fock
representation accounts for the entire memory space.

\subsection{Canonical coordinates and finite thermal factors}
\label{subsec:finite-thermal}

For each $\omega>0$, choose orthonormal pairs
$e_{1,\ell},-J_{1,\omega}e_{1,\ell}$ in
$E_{1,\omega}$ and transport them to $E_{2,\omega}$
through the isometry $\mathcal S$ of
Proposition~\ref{prop:alignment}. Let $\mathcal I$
be the resulting finite or countable set of mode
indices, with frequency $\omega_\ell$ assigned to
each $\ell\in\mathcal I$. Define
\begin{equation}\label{eq:canonical-coordinates}
\mathfrak q_{i,\ell}
:=\kappa_{\omega_\ell}^{-1/2}e_{i,\ell},
\qquad
\mathfrak p_{i,\ell}
:=-\kappa_{\omega_\ell}^{-1/2}
J_{i,\omega_\ell}e_{i,\ell},
\qquad
q_\ell:=e^{-\omega_\ell}.
\end{equation}
By \eqref{eq:balance}, these vectors satisfy
$\sigma_{E_i}(\mathfrak q_{i,\ell},
\mathfrak p_{i,m})=\delta_{\ell m}$, with all
other canonical pairings vanishing.

The corresponding self-adjoint memory fields
$Q_{i,\ell}:=F_{\mathfrak q_{i,\ell}}$ and
$P_{i,\ell}:=F_{\mathfrak p_{i,\ell}}$
transform under modular evolution as
\begin{equation}\label{eq:rotation}
\begin{aligned}
\alpha_u^{(i)}(Q_{i,\ell})
&=\cos(\omega_\ell u)Q_{i,\ell}
-\sin(\omega_\ell u)P_{i,\ell},\\
\alpha_u^{(i)}(P_{i,\ell})
&=\sin(\omega_\ell u)Q_{i,\ell}
+\cos(\omega_\ell u)P_{i,\ell}.
\end{aligned}
\end{equation}
These rotations agree with those generated by the
one-mode thermal state $\tau(q_\ell)$ , where
$q_\ell=e^{-\omega_\ell}$.

The identification \eqref{eq:alignment} also gives
common coordinates for the projected physical
directions. Let $\mathcal Y:=\ell^2(\mathcal I;\R^2)$ and define
the coordinate map $L$ by
\begin{equation}\label{eq:coupling}
P_{i,\mathrm{rot}}\iota_i s
=\sum_{\ell\in\mathcal I}
\left((Ls)_{\ell,1}\mathfrak q_{i,\ell}
+(Ls)_{\ell,2}\mathfrak p_{i,\ell}\right),
\qquad s\in\R^d.
\end{equation}
The coefficients are independent of $i$, since
$\mathcal S$ identifies the corresponding canonical
bases and projected physical directions.

For finite-support $f\in\mathcal Y$, orthogonality
of the canonical coordinates gives
\begin{equation}\label{eq:coordinate-norm}
\left\|\sum_\ell
(f_{\ell,1}\mathfrak q_{i,\ell}
+f_{\ell,2}\mathfrak p_{i,\ell})
\right\|_{E_i}^2
=\frac12\sum_\ell v_\ell|f_\ell|^2
\ge\frac12\|f\|_{\mathcal Y}^2,
\end{equation}
where $v_\ell:=(1+q_\ell)/(1-q_\ell)$.
The estimate extends to convergent coordinate
expansions and gives
$\|Ls\|_{\mathcal Y}
\le\sqrt2\,\|\iota_i s\|_{E_i}$.
Thus $L:\R^d\to\mathcal Y$ is bounded.

We now show that these canonical modes are realized
as tensor factors of the original memories.
Since the argument is the same for both inputs,
we fix either input and omit the index $i$.

\begin{lemma}[Finite-mode factorization]
\label{lem:finite-factor}
For every finite $F\subset\mathcal I$, there
is a unitary memory identification under which
\begin{equation}\label{eq:finite-factor}
\cH_U\simeq\cF(\C^F)\otimes\cH_{R_F},
\qquad
\mu=\tau_F\otimes\nu_F,
\qquad
\tau_F:=\bigotimes_{\ell\in F}\tau(q_\ell),
\end{equation}
where $\nu_F$ is a faithful normal state.
\end{lemma}

\begin{proof}

For a finite set $F\subset\mathcal I$, the canonical
fields $Q_\ell,P_\ell$, $\ell\in F$, generate a regular
Weyl representation with nondegenerate symplectic form.
By the Stone--von Neumann theorem with multiplicity
\cite[Theorem~4]{Derezinski}, there is a unitary
identification
\[
\cH_U\simeq\cF(\C^F)\otimes\cH_{R_F},
\]
under which the Weyl operators of the selected modes
act on the Fock factor and trivially on $\cH_{R_F}$.
By irreducibility of the Fock representation
\cite[Theorem~3.10]{PetzCCR}, the von Neumann algebra
generated by these modes is
\[
\mathcal N_F
=\cB(\cF(\C^F))\otimes I_{R_F}.
\]

By \eqref{eq:rotation}, the modular evolution
preserves the algebra $\mathcal N_F$, that is,
$\alpha_u(\mathcal N_F)=\mathcal N_F$ for all
$u\in\R$. Takesaki's theorem \cite{Takesaki}
states that a von Neumann subalgebra invariant
under the modular group of a faithful normal state
admits a normal, state-preserving conditional
expectation. Hence there exists
$\mathcal E_F:\cB(\cH_U)\to\mathcal N_F$
satisfying
\[
\Tr[\mu\mathcal E_F(Z)]=\Tr(\mu Z),
\qquad
\mathcal E_F(XZY)=X\mathcal E_F(Z)Y,
\]
for $Z\in\cB(\cH_U)$ and $X,Y\in\mathcal N_F$,
where the second identity is the bimodularity
property.

For $X\in\mathcal N_F$ and
$Y\in\cB(\cH_{R_F})$, bimodularity gives
\[
X\mathcal E_F(I\otimes Y)
=\mathcal E_F(X(I\otimes Y))
=\mathcal E_F((I\otimes Y)X)
=\mathcal E_F(I\otimes Y)X.
\]
Thus $\mathcal E_F(I\otimes Y)$ lies in the
center of $\mathcal N_F$, which is $\C I$.
Hence $\mathcal E_F(I\otimes Y)=\varphi_F(Y)I$,
where $\varphi_F$ is a normal state by the positivity,
unitality, and normality of $\mathcal E_F$.
Since $\mathcal E_F$ preserves $\mu$, we obtain
$\Tr[\mu(X\otimes Y)] =\Tr[\mu\mathcal E_F(X\otimes Y)] =\Tr[\mu(X\otimes I)\mathcal E_F(I\otimes Y)]
=\Tr[\mu(X\otimes I)]\,\varphi_F(Y)$.

Let $\mu_F:=\Tr_{R_F}\mu$ and
$\varphi_F(Y)=\Tr(\nu_FY)$ for a normal state
$\nu_F$. Then
\[
\Tr[\mu(X\otimes Y)]
=\Tr(\mu_FX)\Tr(\nu_FY)
=\Tr[(\mu_F\otimes\nu_F)(X\otimes Y)].
\]

Since finite linear combinations of elementary tensors
are ultraweakly dense, the preceding identity implies
$\mu=\mu_F\otimes\nu_F$.
Faithfulness of $\mu$ implies that both factors
are faithful.

By \eqref{eq:rotation}, the modular automorphism
groups of $\mu_F$ and $\tau_F$ coincide on the Weyl
generators, and hence on $\cB(\cF(\C^F))$.
Consequently,
$\tau_F^{-\ii u}\mu_F^{\ii u}$ is scalar for every
$u\in\R$. It follows that $\mu_F$ and $\tau_F$
are proportional. Since both are normalized,
$\mu_F=\tau_F$, proving \eqref{eq:finite-factor}.
\end{proof}

\subsection{The countable Fock factor and residual memory}
\label{subsec:countable-fock}

We extend the finite-mode factorization
\eqref{eq:finite-factor} to all modes in $\mathcal I$.
The main step is to show that the mutually orthogonal
occupation subspaces associated with the canonical
modes span the entire memory Hilbert space.

\begin{proposition}[Normal thermal realization]
\label{prop:fock}
The canonical modes in \eqref{eq:canonical-coordinates}
admit unitary memory identifications
\begin{equation}\label{eq:fock-factor}
\cH_{U_i}\simeq\cH_Q\otimes\cH_{K_i},
\qquad
\mu_i=\tau_Q\otimes\nu_i,
\qquad
\cH_Q:=\cF(\ell^2(\mathcal I;\C)),
\end{equation}
where $\nu_i$ is faithful and
$\tau_Q:=\bigotimes_{\ell\in\mathcal I}\tau(q_\ell)$
is a normal product thermal state satisfying
$\sum_\ell q_\ell<\infty$ and
$S(\tau_Q)\le S(\mu_i)<\infty$.

Under these identifications,
\begin{equation}\label{eq:weyl-factor}
V_i(s)=W_Q(Ls)\otimes Z_i(s),
\qquad [Z_i(s),\nu_i]=0,
\end{equation}
where $W_Q$ is the standard Weyl representation
on $\cH_Q$ and
$Z_i:\R^d\to\mathcal U(\cH_{K_i})$
is a strongly continuous commuting unitary
representation.
\end{proposition}

\begin{proof}
By Lemma~\ref{lem:finite-factor}, for every finite
$F\subset\mathcal I$,
$\|\mu_i\|=\|\tau_F\|\,\|\nu_{i,F}\|
\le\prod_{\ell\in F}(1-q_\ell)$,
since $\|\nu_{i,F}\|\le1$ and
$\|\tau(q_\ell)\|=1-q_\ell$.
Taking the limit over finite $F$ gives
\begin{equation}\label{eq:vacuum-weight}
c:=\prod_{\ell\in\mathcal I}(1-q_\ell)
\ge\|\mu_i\|>0.
\end{equation}
Since $-\log(1-q_\ell)\ge q_\ell$,
it follows that $\sum_\ell q_\ell<\infty$.

For each mode $\ell$, let $\Pi_{i,\ell,0}$
denote its vacuum projection, and define
$\Pi_{i,0}:=\bigwedge_{\ell\in\mathcal I}
\Pi_{i,\ell,0}$.
For finite $F\subset\mathcal I$, the projection
$\Pi_{i,F,0}:=\prod_{\ell\in F}\Pi_{i,\ell,0}$
satisfies
$\Tr(\mu_i\Pi_{i,F,0})
=\prod_{\ell\in F}(1-q_\ell)$
by \eqref{eq:finite-factor}.
Since $\Pi_{i,F,0}$ decreases strongly to
$\Pi_{i,0}$, normality gives
$\Tr(\mu_i\Pi_{i,0})=c>0$.

For each mode $\ell$, let $E_{i,\ell}(r,s)$
denote the occupation matrix units associated
with its canonical Weyl representation.
Their compatibility with the finite-mode
factorizations and their commutation for
distinct modes are established in
Appendix~\ref{app:occupation:matrix-units}.
Their products define mutually orthogonal
occupation projections $\Pi_{i,\mathbf n}$,
indexed by finite-support sequences
$\mathbf n=(n_\ell)_{\ell\in\mathcal I}
\in\N_0^{(\mathcal I)}$.

The modular eigenoperator relations derived in
Appendix~\ref{app:occupation:weights} give
\begin{equation}\label{eq:occupation-weights}
\Tr(\mu_i\Pi_{i,\mathbf n})
=cq^{\mathbf n},
\qquad
q^{\mathbf n}:=\prod_{\ell\in\mathcal I}
q_\ell^{n_\ell}.
\end{equation}

Set $\cH_{K_i}:=\Pi_{i,0}\cH_{U_i}$.
The occupation matrix units define an isometry
$\mathcal U_i:\cH_Q\otimes\cH_{K_i}\to\cH_{U_i}$
whose range is the orthogonal sum of
$\Ran\Pi_{i,\mathbf n}$ over finite-support
occupation sequences. The construction of
this isometry and the corresponding density
matrix-element identities are given in
Appendix~\ref{app:occupation:factorization}.

To establish that $\mathcal U_i$ is onto,
we sum the occupation weights. By
\eqref{eq:vacuum-weight} and
\eqref{eq:occupation-weights},
\begin{equation}\label{eq:exhaustion}
\sum_{\mathbf n\in\N_0^{(\mathcal I)}}
\Tr(\mu_i\Pi_{i,\mathbf n})
=c\prod_{\ell\in\mathcal I}(1-q_\ell)^{-1}=1.
\end{equation}
Since the projections $\Pi_{i,\mathbf n}$
are mutually orthogonal, their sum is the
projection onto $\Ran\mathcal U_i$.
By normality, \eqref{eq:exhaustion} implies
that the orthogonal complement of this range
has zero $\mu_i$-weight. Faithfulness of
$\mu_i$ therefore implies that the complement
vanishes, so $\mathcal U_i$ is unitary.

The matrix-element identity
\eqref{eq:occupation-density-blocks} shows that
the transformed density has no off-diagonal
occupation blocks, while each diagonal block
equals $q^{\mathbf n}$ times the vacuum density.
Normalizing the latter by $c$ gives
$\mu_i=\tau_Q\otimes\nu_i$, with $\nu_i$ faithful.

Entropy additivity and monotone convergence
over finite thermal factors yield
$S(\tau_Q)=\sum_\ell S(\tau(q_\ell))
\le S(\mu_i)<\infty$.

It remains to identify $V_i(s)$.
By Lemma~\ref{lem:balance}, the fixed fields
commute with all rotating fields and are
invariant under modular evolution.
Since the rotating Weyl operators generate
$\cB(\cH_Q)\otimes I$, their commutant is
$I\otimes\cB(\cH_{K_i})$.
The fixed part of \eqref{eq:field-weyl},
including the mean phase, therefore defines
$Z_i(s)$ through
\[
e^{\ii b_{0,i}^{\mathsf T}s}
\mathscr W_i(P_{i,0}\iota_i s)
=I\otimes Z_i(s).
\]
Since $\sigma_{E_i}$ vanishes on $E_{i,0}$,
the Weyl relations give
$Z_i(s)Z_i(t)=Z_i(s+t)$ and $Z_i(0)=I$.
Thus $Z_i$ is a strongly continuous
commuting unitary representation of $\R^d$.
Its modular invariance gives
$[Z_i(s),\nu_i]=0$.

By \eqref{eq:coupling}, the coefficients $Ls$
describe the rotating part of $\iota_i s$
in the canonical position--momentum basis.
Under the Fock identification, its Weyl
operator is
$\mathscr W_i(P_{i,\mathrm{rot}}\iota_i s)
=W_Q(Ls)\otimes I$.
Combining the rotating and fixed parts gives
\eqref{eq:weyl-factor}.

Finally, $\sum_\ell q_\ell<\infty$ implies
$\sup_\ell v_\ell<\infty$, so the weighted
norm in \eqref{eq:coordinate-norm} is equivalent
to the standard norm on $\mathcal Y$.
Hence the factorization is compatible with
the completed Weyl representation.
\end{proof}

The two memories therefore share the thermal
state $\tau_Q$ and Weyl coupling $L$, while
their residual representations $Z_i$ may differ.
We next use this structure to reconstruct
the original joint states.

\section{Joint-state reconstruction and converse}
\label{sec:reconstruction}

\subsection{Orthogonal displacement sectors and Gaussian reconstruction}
\label{subsec:gaussian-reconstruction}
We now recover the joint states from the factorization
\eqref{eq:weyl-factor}. Since $\nu_i$ is faithful and trace class,
its eigenspaces are finite-dimensional. They are invariant under
every $Z_i(s)$ because $[Z_i(s),\nu_i]=0$. The operators $Z_i(s)$
commute, so they can be simultaneously diagonalized on each
eigenspace. Grouping the resulting vectors by their characters
gives
\begin{equation}\label{eq:residual-sectors}
\begin{aligned}
\cH_{K_i}
&=\bigoplus_{j\in\mathcal J_i}\cH_{K_j^{(i)}},\qquad
Z_i(s)
=\bigoplus_j e^{\ii(b_j^{(i)})^{\mathsf T}s}I_{K_j^{(i)}},
\quad
\nu_i=\bigoplus_j p_j^{(i)}\nu_j^{(i)}.
\end{aligned}
\end{equation}
Here $b_j^{(i)}\in\R^d$, the weights $p_j^{(i)}>0$ sum to one,
and each $\nu_j^{(i)}$ is a faithful density operator.
The decomposition is countable because each eigenspace of
$\nu_i$ is finite-dimensional. Here $\cH_{K_j^{(i)}}$ is the residual memory subspace carrying the character $s\mapsto e^{\ii(b_j^{(i)})^{\mathsf T}s}$; the corresponding full memory sector is
$\cH_Q\otimes\cH_{K_j^{(i)}}$, as in
Theorem~\ref{thm:main}.

Using \eqref{eq:classification}, \eqref{eq:fock-factor},
and \eqref{eq:residual-sectors} in the definition of the
partial characteristic function, we obtain
\begin{equation}\label{eq:block-coefficients}
\Phi_i(s)=\bigoplus_j p_j^{(i)}
e^{-s^{\mathsf T}Gs/4+\ii(b_j^{(i)})^{\mathsf T}s}
\tau_Q^{1/2}W_Q(Ls)\tau_Q^{1/2}\otimes\nu_j^{(i)}.
\end{equation}

Let $P_j^{(i)}$ be the projection of
$\cH_{K_i}$ onto $\cH_{K_j^{(i)}}$, and,
under the memory identification above, define
\[
\begin{aligned}
R_{ij}
&:=(I_{AQ}\otimes P_j^{(i)})\rho_i
   (I_{AQ}\otimes P_j^{(i)}),\qquad
\eta_j^{(i)}
:=\frac{1}{p_j^{(i)}}
  \Tr_{K_j^{(i)}}R_{ij}.
\end{aligned}
\]
The zero-argument case of
\eqref{eq:block-coefficients} gives
$\Tr R_{ij}=p_j^{(i)}>0$.
Thus each $\eta_j^{(i)}$ is a positive
normal state on $AQ$, with partial
characteristic function
\[
\Tr_A[\eta_j^{(i)}(W_A(s)\otimes I_Q)]
=e^{-s^{\mathsf T}Gs/4+
\ii(b_j^{(i)})^{\mathsf T}s}
\tau_Q^{1/2}W_Q(Ls)\tau_Q^{1/2}.
\]

Equation~\eqref{eq:block-coefficients}
shows that the operator-valued Weyl
coefficients of $R_{ij}$ coincide with
those of
$p_j^{(i)}\eta_j^{(i)}\otimes\nu_j^{(i)}$.
Since the physical Weyl operators generate
$\cB(\cH_A)$ as a von Neumann algebra,
uniqueness of normal Weyl coefficients gives
$
R_{ij}
=p_j^{(i)}\eta_j^{(i)}\otimes\nu_j^{(i)}.
$
The off-diagonal memory blocks have vanishing
Weyl coefficients by the same identity and
therefore vanish. Consequently,
\[
\rho_i\simeq
\bigoplus_{j\in\mathcal J_i}
p_j^{(i)}\eta_j^{(i)}\otimes\nu_j^{(i)}.
\]

To identify the states $\eta_j^{(i)}$, define
the bounded operators on $\mathcal Y$
\begin{equation}\label{eq:thermal-covariances}
\mathsf V:=\bigoplus_\ell v_\ell I_2,\qquad
\mathsf C:=\bigoplus_\ell c_\ell I_2,
\qquad
v_\ell:=\frac{1+q_\ell}{1-q_\ell},
\quad
c_\ell:=\frac{2\sqrt{q_\ell}}{1-q_\ell}.
\end{equation}
Here $\mathsf V$ is the covariance of
the thermal memory $\tau_Q$, while
$\mathsf C$ determines its covariance
with the physical system. Both are
bounded because $\sum_\ell q_\ell<\infty$.

For $s\in\R^d$ and finite-support
$f\in\mathcal Y$, we have
\[
\begin{aligned}
\chi_j^{(i)}(s,f)
&:=\Tr[\eta_j^{(i)}
(W_A(s)\otimes W_Q(f))]
=e^{-s^{\mathsf T}Gs/4+
\ii(b_j^{(i)})^{\mathsf T}s}
\Tr[\tau_Q^{1/2}W_Q(Ls)
\tau_Q^{1/2}W_Q(f)].
\end{aligned}
\]
The thermal sandwich identity
\begin{equation}\label{eq:sandwich}
\Tr[\tau_Q^{1/2}W_Q(a)
\tau_Q^{1/2}W_Q(f)]
=\exp\left[
-\tfrac14\ip a{\mathsf Va}
-\tfrac12\ip a{\mathsf Cf}
-\tfrac14\ip f{\mathsf Vf}
\right],
\end{equation}
proved in Appendix~\ref{app:sandwich},
then gives
\begin{equation}\label{eq:joint-characteristic}
\begin{aligned}
\chi_j^{(i)}(s,f)=\exp\Bigl[
&\ii(b_j^{(i)})^{\mathsf T}s
-\tfrac14s^{\mathsf T}
(G+L^{\mathsf T}\mathsf VL)s-\tfrac12\ip{Ls}{\mathsf Cf}
-\tfrac14\ip f{\mathsf Vf}
\Bigr],
\end{aligned}
\end{equation}
where $L^{\mathsf T}:\mathcal Y\to\R^d$
is the real Hilbert-space adjoint.

Thus each $\eta_j^{(i)}$ is Gaussian.
For a finite set $F\subset\mathcal I$,
let $L_F$ denote the restriction of $L$
to the corresponding coordinates, and
let $\mathsf V_F,\mathsf C_F$ be the
associated diagonal blocks. The covariance
of the reduced state on $A$ and these
memory modes is
\begin{equation}\label{eq:joint-covariance}
\Gamma_F=
\begin{pmatrix}
G+L^{\mathsf T}\mathsf VL
&L_F^{\mathsf T}\mathsf C_F\\
\mathsf C_FL_F&\mathsf V_F
\end{pmatrix}.
\end{equation}
The physical covariance retains the
contribution of all memory modes, including
those outside $F$. Since $\Gamma_F$ is the
covariance of a positive state, it satisfies
the quantum uncertainty condition.

These covariance matrices are independent
of $i$ and $j$: $G$ is common by
Proposition~\ref{prop:classification},
while the thermal parameters and $L$
are common by
Proposition~\ref{prop:alignment}
and \eqref{eq:coupling}.
Moreover, normal states on $AQ$ are
determined by their finite-mode Weyl
characteristic functions. Hence all
$\eta_j^{(i)}$ have the same covariance
and differ only by their physical
displacements $b_j^{(i)}$.

Let $\gamma_{AQ}$ be the centered Gaussian
state obtained by removing the displacement
from any one block. By
\eqref{eq:joint-characteristic} and Weyl
uniqueness,
$\eta_j^{(i)}=\gamma_{AQ}^{b_j^{(i)}}$
for every $i,j$. Therefore,
\[
\rho_i\simeq
\bigoplus_{j\in\mathcal J_i}
p_j^{(i)}
\gamma_{AQ}^{b_j^{(i)}}
\otimes\nu_j^{(i)},
\]
as asserted in Theorem~\ref{thm:main}.

Finally, the memory factorization and
the direct-sum entropy formula give the
stated entropy identities. Since
$\gamma_A$ is centered, the physical
energy identity follows from
$\Tr(\gamma_A^bH_A)
=\Tr(\gamma_AH_A)+|b|^2/2$.
The assumptions \eqref{eq:assumptions}
ensure the required finiteness conditions.

\subsection{Balanced mixing of the common Gaussian core}
\label{subsec:converse}

Assume the normal form in Theorem~\ref{thm:main}, and fix the residual-memory indices

$j\in\mathcal J_1$ and $k\in\mathcal J_2$.
In addition to the physical beam splitter,
apply a balanced beam splitter to the two
memory factors $Q_1,Q_2$ as well. This unitary
mixes their Weyl coordinates by
$(f,g)\mapsto((f+g)/\sqrt2,(f-g)/\sqrt2)$
and is well defined also for countably many
modes. Since it acts only on the conditioning
memory, it leaves conditional mutual
information unchanged.

The centered Gaussian characteristic function
$\chi_\gamma(x)$ satisfies
\begin{equation}\label{eq:gaussian-parallelogram}
\chi_\gamma((x+y)/\sqrt2)
\chi_\gamma((x-y)/\sqrt2)
=\chi_\gamma(x)\chi_\gamma(y),
\end{equation}
where $x,y$ contain physical coordinates
and finite-support memory coordinates.
Thus the jointly mixed Gaussian cores factor
across $(A_+Q_+):(A_-Q_-)$.

The physical displacements become
$b_{+,jk}:=(b_j^{(1)}+b_k^{(2)})/\sqrt2$
and
$b_{-,jk}:=(b_j^{(1)}-b_k^{(2)})/\sqrt2$.
They act locally on the respective physical
outputs and do not change conditional mutual
information. The residual memories remain
independent, while the orthogonal pair of
labels $(j,k)$ lies entirely in the conditioning
memory.

Denote by $\widetilde\omega^{jk}$ the normalized output
state conditional on the residual-memory indices
$(j,k)$ after the memory-only beam splitter. The factorization just established gives
$I(A_+:A_-\mid Q_+Q_-K_j^{(1)}K_k^{(2)})_{
\widetilde\omega^{jk}}=0$ for every pair. Since the
orthogonal labels belong to the conditioning memory,
the direct-sum formula and invariance under its unitary
transformation yield
\[
 \begin{aligned}
 I(A_+:A_-\mid U_1U_2)_\omega
 &=\sum_{j,k}p_j^{(1)}p_k^{(2)}
 I(A_+:A_-\mid Q_+Q_-K_j^{(1)}K_k^{(2)})_{
 \widetilde\omega^{jk}}=0.
 \end{aligned}
\]
Thus averaging over the orthogonal residual-memory
subspaces does not introduce conditional correlations
between the physical outputs.

For countably many modes, the factorization
in \eqref{eq:gaussian-parallelogram} extends
from finite-mode Weyl operators by normal
Weyl uniqueness. This proves the converse
and completes Theorem~\ref{thm:main}.

\section{Conclusion}
\label{sec:conclusion}

We have established a conditional quantum
Kac--Bernstein theorem for two independent,
not necessarily identical, bosonic inputs with
quantum memories. We showed that conditional
independence of the beam-splitter outputs is
equivalent to a common Gaussian structure,
up to physical displacements and orthogonal
residual-memory sectors. The common bosonic
memory factor has a product thermal marginal,
while the residual memory states may differ.

The proof combines an operator Bernstein
equation with modular commutation relations
and a thermal Fock-space factorization.
In particular, when the common bosonic
memory factor is trivial, both inputs are
separable across the physical system and
memory. Thus entangled equality states
require infinite-dimensional bosonic quantum memories.

Our result extends the classical conditional
Gaussian characterization to quantum side
information and recovers the unconditioned
quantum Kac--Bernstein theorem as a special case.

\section*{Acknowledgments}

The authors thank Chandra Nair for bringing this
problem to their attention, pointing out the absence
of a conditional quantum Kac--Bernstein theorem
in the existing literature, and sharing valuable
insights and intuition that helped shape this work.

The work of H. Hosseini and J. Zhao was supported by the Hong Kong Research Grants Council (RGC) under Grant Nos. 14213826 and 14309924, and by the 1+1+1 CUHK--CUHK(SZ)--GDST Joint Collaboration Fund (Grant No. GRDP2025-022).

The work of M. Hayashi was supported in part by the General R\&D Projects of the 1+1+1 CUHK--CUHK(SZ)--GDST Joint Collaboration Fund (Grant No. GRDP2025-022) and the Guangdong Provincial Quantum Science Strategic Initiative (Grant No. GDZX2505003).

\section*{Declaration of AI Assistance}

The authors used OpenAI's GPT-5.6 and GPT-6
models in thinking mode during the preparation
of this manuscript. AI assistance contributed
to extending the initial finite-dimensional
approach to the infinite-dimensional setting.
In particular, the use of canonical commutation
relations and Weyl representations in the modular
analysis and thermal Fock-space construction
(Sections~\ref{sec:geometry} and
\ref{sec:thermal}) was suggested through AI-assisted
discussions. AI was also used to refine
mathematical arguments, examine technical details,
and improve the completeness of the proofs.
Additional assistance included language editing,
organization, and presentation of the manuscript.

All AI-assisted mathematical arguments and written
material were critically reviewed, verified, and
revised by the authors, who take full responsibility
for the accuracy and integrity of the manuscript.

\clearpage
\appendix
\begin{center}
    \Large\bfseries Appendices
\end{center}

\section{Markov structure on the marginal support}
\label{app:markov}

We prove Lemma~\ref{lem:markov-support}, extending the
faithful-state Markov decomposition to states that need
not be jointly faithful. The key observation is that local, injective regularization
can make the joint state faithful without changing its
memory marginal or destroying conditional independence.
The resulting Markov decomposition can then be transferred
back to the original state by injectivity.

\begin{proof}[Proof of Lemma~\ref{lem:markov-support}]
We restrict $M$ to $\supp\zeta_M$ and choose faithful
finite-entropy states $\vartheta_B,\vartheta_C$.
For $D=B,C$ and $0<\varepsilon<1$, define
\[
\mathcal R_D(X)
:=\varepsilon X+(1-\varepsilon)\vartheta_D\Tr X.
\]
These channels are injective on arbitrary auxiliary
extensions. Indeed, their inverses on the respective
ranges are given by
\begin{equation}\label{eq:regularization-inverse}
Z\longmapsto
\varepsilon^{-1}
\bigl[Z-(1-\varepsilon)
\vartheta_D\otimes\Tr_D Z\bigr],
\end{equation}
with the tensor factors in their original order.

Set
\[
\widetilde\zeta
:=(\mathcal R_B\otimes\id_M\otimes\mathcal R_C)(\zeta).
\]
Since
\[
\widetilde\zeta
\ge (1-\varepsilon)^2
\vartheta_B\otimes\zeta_M\otimes\vartheta_C,
\]
the state $\widetilde\zeta$ is faithful. All its
relevant entropies are finite, by the finite-mixture
entropy bound and subadditivity. Moreover, endpoint
data processing gives
\[
0\le I(B:C\mid M)_{\widetilde\zeta}
\le I(B:C\mid M)_\zeta=0.
\]
The faithful Markov decomposition
\cite[Section~6, Theorem~8]{JP} therefore yields
a unitary identification
\[
V_M:\supp\zeta_M
\longrightarrow
\bigoplus_a\cH_{L_a}\otimes\cH_{R_a}
\]
under which
\[
\widetilde\zeta
=\bigoplus_a r_a\,
\widetilde\eta^a_{BL_a}
\otimes\widetilde\theta^a_{R_aC}.
\]
Here and below, states are expressed in this
identified representation. Since the endpoint
channels preserve the memory marginal, the same
unitary identifies $\zeta_M$.

Let $P_a$ denote the corresponding memory
projections. The endpoint channels commute with
memory compression, and their tensor product is
injective. Hence
\[
P_a\widetilde\zeta P_b=0
\quad (a\ne b)
\qquad\Longrightarrow\qquad
P_a\zeta P_b=0.
\]
Consequently,
\[
\zeta=\bigoplus_a r_a\zeta^a,
\]
where each $\zeta^a$ is a normalized positive state.

Let $\eta^a_{BL_a}$ and $\theta^a_{R_aC}$
be the corresponding marginals of $\zeta^a$.
Because the endpoint channels act locally,
\[
(\mathcal R_B\otimes\id_{L_a})(\eta^a)
=\widetilde\eta^a,
\qquad
(\id_{R_a}\otimes\mathcal R_C)(\theta^a)
=\widetilde\theta^a.
\]
It follows that $\zeta^a$ and
$\eta^a_{BL_a}\otimes\theta^a_{R_aC}$
have the same image under the injective
regularization map. Thus
\[
\zeta^a
=\eta^a_{BL_a}\otimes\theta^a_{R_aC},
\]
which establishes the claimed decomposition.
\end{proof}

\section{Weighted fields, domains, and completion}
\label{app:fields}

We prove Lemma~\ref{lem:fields}. As in
Section~\ref{subsec:fields}, we fix one input and
omit its index. The construction proceeds from the
second-moment bounds to the Weyl representation on
$\mathcal D$, and then extends that representation
to the completed covariance space $E$.

\subsection{Transfer of second moments}
\label{app:fields:moments}

For $V(te_k)=e^{\ii tA_k}$, set
\[
\psi_k(t):=\Tr(\mu e^{\ii tA_k}),\qquad
\phi_k(t):=\Tr(\rho_AW_A(te_k)).
\]
Taking the trace in \eqref{eq:normalization}
and applying \eqref{eq:classification} gives
\begin{equation}\label{eq:moment-characteristic}
\psi_k(t)=e^{G_{kk}t^2/4}\phi_k(t).
\end{equation}
Since $m_{2,k}:=\Tr(\rho_AR_{A,k}^2)<\infty$,
the spectral theorem and dominated convergence yield
\[
\lim_{t\to0}
\frac{2(1-\Rea\phi_k(t))}{t^2}=m_{2,k},
\]
using $2(1-\cos(tx))/t^2\le x^2$. Hence
\begin{equation}\label{eq:curvature}
\lim_{t\to0}
\frac{2(1-\Rea\psi_k(t))}{t^2}
=m_{2,k}-\tfrac12G_{kk}.
\end{equation}

Let $M_k(B):=\Tr(\mu E_{A_k}(B))$ be the spectral
probability measure of $A_k$ in the state $\mu$.
Then
\[
\frac{2(1-\Rea\psi_k(t))}{t^2}
=\int_{\R}\frac{2(1-\cos(tx))}{t^2}\,dM_k(x).
\]
Fatou's lemma and \eqref{eq:curvature} first imply
$\int x^2\,dM_k(x)<\infty$. Dominated convergence
then identifies this second moment with the limit
in \eqref{eq:curvature}. Differentiating
\eqref{eq:moment-characteristic} at zero, now
justified by the finite first moments, proves
\eqref{eq:moment-transfer}.

In particular, the centered fields
$B_{k,u}$ of \eqref{eq:centered-fields} satisfy
\[
\Tr(\mu B_{k,u})=0,\qquad
\Tr(\mu B_{k,u}^2)<\infty.
\]
Centering changes only scalar phases, so their
commutation relations remain those of
\eqref{eq:translated-ccr}.

\subsection{Weyl products and generator domains}
\label{app:fields:domains}

Let $a=\sum_{j=1}^m c_j\varepsilon_j\in\mathcal D$,
where $\varepsilon_j=\varepsilon_{k_j,u_j}$.
Write $B_j:=B_{k_j,u_j}$ and
$\sigma_{jk}:=\sigma(\varepsilon_j,\varepsilon_k)$.
We define
\begin{equation}\label{eq:ordered-weyl}
\mathscr W(a):=
\exp\left(\frac{\ii}{2}
\sum_{j<k}c_jc_k\sigma_{jk}\right)
\prod_{j=1}^m e^{\ii c_jB_j}.
\end{equation}

To see the role of the scalar phase, consider

$a=c_1\varepsilon_1+c_2\varepsilon_2$.
The commutation relation gives
\[
e^{\ii c_1B_1}e^{\ii c_2B_2}
=e^{-\ii c_1c_2\sigma_{12}}
e^{\ii c_2B_2}e^{\ii c_1B_1},
\]
so
\[
\mathscr W(a)
=e^{-\ii c_1c_2\sigma_{12}/2}
e^{\ii c_2B_2}e^{\ii c_1B_1}.
\]

This is exactly the expression obtained by
reversing the order of the two terms.
The same argument applies to adjacent factors
in the general product, showing that
$\mathscr W(a)$ is independent of their ordering.
Together with the one-parameter group laws,
it also gives
\begin{equation}\label{eq:free-weyl}
\mathscr W(a)\mathscr W(b)
=e^{-\ii\sigma(a,b)/2}\mathscr W(a+b),
\qquad
\mathscr W(a)^*=\mathscr W(-a).
\end{equation}

The product formula has two consequences that will
be used separately: it produces a self-adjoint
generator for each finite formal sum, and it gives
the Weyl relation required for the later completion.

For each $a\in\mathcal D$, the map
$t\mapsto\mathscr W(ta)$ is a strongly
continuous one-parameter unitary group.
By Stone's theorem, it has a self-adjoint
generator $F_a$ satisfying
$\mathscr W(ta)=e^{\ii tF_a}$.

To pass from the generator on eigenvectors to
the weighted operator $F_a\mu^{1/2}$, we first
establish a common domain and then use closedness.

Diagonalize $\mu e_r=u_re_r$, where $u_r>0$,
and set $\mathcal D_0:=\operatorname{span}\{e_r\}$.
The second-moment bounds imply
$\mathcal D_0\subset\Dom B_{k,u}$ for every $k,u$.

For $U_j(t):=e^{\ii t c_jB_j}$ and
$\xi\in\mathcal D_0$, the telescoping identity
\[
\left(\prod_{j=1}^mU_j(t)-I\right)\xi
=\sum_{j=1}^m
U_1(t)\cdots U_{j-1}(t)(U_j(t)-I)\xi
\]
gives the derivative
$\ii\sum_jc_jB_j\xi$ at zero.
The phase in \eqref{eq:ordered-weyl} for $ta$
is $1+O(t^2)$, so Stone's domain characterization \cite[Section~3.3]{Derezinski}
yields
\begin{equation}\label{eq:generator-sum}
\xi\in\Dom F_a,\qquad
F_a\xi=\sum_jc_jB_j\xi.
\end{equation}
Consequently,
$\sum_r u_r\norm{F_ae_r}^2<\infty$.

Define $\Lambda(a)$ by
$\Lambda(a)e_r:=\sqrt{u_r}F_ae_r$.
This is Hilbert--Schmidt. If $P_N$ projects onto
the first $N$ eigenvectors of $\mu$, then
\[
\mu^{1/2}P_N\xi\to\mu^{1/2}\xi,\qquad
F_a\mu^{1/2}P_N\xi
=\Lambda(a)P_N\xi\to\Lambda(a)\xi.
\]
Closedness of $F_a$ therefore gives
\begin{equation}\label{eq:weighted-domain}
\Ran\mu^{1/2}\subset\Dom F_a,\qquad
F_a\mu^{1/2}=\Lambda(a).
\end{equation}
By \eqref{eq:generator-sum}, the map
$a\mapsto\Lambda(a)$ is real-linear. Moreover,
\[
\Tr(\mu F_a)
=\ip{\mu^{1/2}}{\Lambda(a)}_{\HS}=0.
\]

\subsection{Covariance and the quotient space}
\label{app:fields:covariance}

The spectral theorem and \eqref{eq:weighted-domain}
give the Hilbert--Schmidt derivative
\begin{equation}\label{eq:weighted-derivative}
\frac{(\mathscr W(ta)-I)\mu^{1/2}}{t}
\longrightarrow\ii\Lambda(a).
\end{equation}
Indeed, the squared error is bounded by the
integrable spectral function $4x^2$.

For $U_s:=\mathscr W(sa)$ and
$V_t:=\mathscr W(tb)$, we have
\[
\Tr[\mu(U_s-I)(V_t-I)]
=\ip{(U_s^*-I)\mu^{1/2}}
{(V_t-I)\mu^{1/2}}_{\HS}.
\]

Subtracting the reversed expression and dividing
by $st$, we apply \eqref{eq:weighted-derivative}.
The resulting limit is

$-2\ii\Ima\ip{\Lambda(a)}{\Lambda(b)}_{\HS}$.
The Weyl relation \eqref{eq:free-weyl} gives $-\ii\sigma(a,b)$ for the same limit. Thus

\begin{equation}\label{eq:weighted-commutator}
\sigma(a,b)
=2\Ima\ip{\Lambda(a)}{\Lambda(b)}_{\HS},
\qquad
|\sigma(a,b)|
\le2\norm{\Lambda(a)}_{\HS}
\norm{\Lambda(b)}_{\HS}.
\end{equation}
In particular, this argument uses bounded
unitary differences rather than products
of unbounded generators.

Recall that
$g(a,b)=\Rea\ip{\Lambda(a)}{\Lambda(b)}_{\HS}$.
If $g(a,a)=0$, then $F_ae_r=0$ for every $r$.
Since $\ker F_a$ is closed and contains the dense
space $\mathcal D_0$, we obtain $F_a=0$ and
$\mathscr W(ta)=I$. Moreover,
\eqref{eq:weighted-commutator} gives
$\sigma(a,b)=0$ for all $b$.

Hence $g$, $\sigma$, and $\mathscr W$ descend
to $\mathcal D/\ker g$. Its Hilbert completion
$E$ is separable, since $\Lambda$ embeds the
quotient isometrically into the separable
Hilbert--Schmidt space.

\subsection{Completion and modular covariance}
\label{app:fields:completion}

We next extend the Weyl operators from the quotient
$\mathcal D/\ker g$ to its completion $E$.
First we obtain strong limits of the bounded unitaries;
only after that do we pass their unbounded generators
to the limit and recover the weighted domain.

For $a\in\mathcal D$, the spectral inequality
$|e^{\ii x}-1|\le|x|$ yields
\[
\norm{(\mathscr W(a)-I)\mu^{1/2}}_{\HS}
\le\norm a_E.
\]
Together with \eqref{eq:free-weyl}, this gives
\begin{equation}\label{eq:weyl-difference}
\norm{(\mathscr W(a)-\mathscr W(b))
\mu^{1/2}}_{\HS}
\le\norm{a-b}_E+
|e^{\ii\sigma(a,b)/2}-1|.
\end{equation}
If $(a_m)$ is Cauchy in $E$, the right-hand
side tends to zero, since
$\sigma(a_m,a_l)=\sigma(a_m-a_l,a_l)$
and \eqref{eq:weighted-commutator} applies.

For bounded $X$,
\[
\norm{Xe_r}\le u_r^{-1/2}
\norm{X\mu^{1/2}}_{\HS}.
\]
Thus $\mathscr W(a_m)$ and $\mathscr W(-a_m)$
converge strongly on $\mathcal D_0$, and hence
on $\cH_U$ by uniform boundedness.
Their limits are adjoints and mutual inverses,
defining a strongly continuous Weyl family on $E$.

It remains to preserve the weighted covariance

under completion. Strong convergence of the unitaries
alone does not identify the domains of their generators;
for that purpose we use their resolvents.
Let $a_m\to a$ in $E$, and write

$F_m:=F_{a_m}$ and
$\Lambda(a_m)\to\Lambda$ in Hilbert--Schmidt norm.
Let $F$ be the generator of $\mathscr W(ta)$.
Strong convergence of the unitary groups implies
strong resolvent convergence, using
\[
(H-\ii I)^{-1}
=\ii\int_0^\infty e^{-t}e^{-\ii tH}\,dt.
\]
Since
$F_me_r\to y_r:=u_r^{-1/2}\Lambda e_r$,
passing to the limit in
\[
e_r=(F_m-\ii I)^{-1}(F_me_r-\ii e_r)
\]
gives $e_r\in\Dom F$ and $Fe_r=y_r$.
Applying the truncation argument of
\eqref{eq:weighted-domain} yields
$F\mu^{1/2}=\Lambda$.
Consequently, the moment identities and
\eqref{eq:field-covariance} remain valid on $E$.

Finally, shifting the modular parameter by $t$
preserves both $g$ and $\sigma$, since
\begin{equation}\label{eq:weighted-modular}
\Lambda(T(t)a)
=\mu^{\ii t}\Lambda(a)\mu^{-\ii t}.
\end{equation}
This defines a strongly continuous orthogonal
group $T(t)$ on $E$, and the modular covariance
of $\mathscr W$ extends from $\mathcal D$ to $E$.

With $\iota e_k:=[\varepsilon_{k,0}]$,
the multiplier in \eqref{eq:classification}
and the definition \eqref{eq:ordered-weyl} give
\[
\mathscr W(\iota s)
=e^{\frac{\ii}{2}\sum_{k<l}s_ks_l\Theta_{kl}}
\prod_{k=1}^d e^{\ii s_k(A_k-b_{0,k}I)}
=e^{-\ii b_0^{\mathsf T}s}V(s).
\]
The modular translates of the symbols
$\iota e_k$ span $\mathcal D$, proving
\eqref{eq:field-orbits} and completing
Lemma~\ref{lem:fields}.

\section{Compatible occupation matrix units}
\label{app:occupation}

We justify the occupation-sector construction used in
Proposition~\ref{prop:fock}. Fix either input and
suppress the index $i$ throughout this appendix.

\subsection{Matrix units of the canonical modes}
\label{app:occupation:matrix-units}

For each $\ell\in\mathcal I$, Stone--von Neumann
with multiplicity gives a normal unital representation
\[
\pi_\ell:\cB(\cF(\C))\longrightarrow\cB(\cH_U)
\]
that maps the standard one-mode Weyl operators to
those of the canonical mode $\ell$. Define
\begin{equation}\label{eq:occupation-matrix-units}
E_\ell(r,s):=\pi_\ell(|r\rangle\langle s|),
\qquad r,s\in\N_0.
\end{equation}
These operators satisfy
$E_\ell(r,s)E_\ell(p,q)=\delta_{sp}E_\ell(r,q)$
and $E_\ell(r,s)^*=E_\ell(s,r)$.

Since the canonical Weyl operators generate
$\cB(\cF(\C))$ as a von Neumann algebra,
the normal representation $\pi_\ell$ is uniquely
determined by its action on them. Hence the matrix
units in \eqref{eq:occupation-matrix-units} are
independent of the finite-mode identification.
Moreover, the Weyl algebras of distinct modes
commute, as do their matrix units. Their finite
products therefore coincide with the occupation
matrix units of the corresponding finite-mode
Fock representation.

The modular rotations \eqref{eq:rotation} are
implemented on each mode by
$q_\ell^{\ii uN_\ell}$, where $N_\ell$
is the number operator. Since
$N_\ell|r\rangle=r|r\rangle$, we obtain
\begin{equation}\label{eq:occupation-modular}
\alpha_u(E_\ell(r,s))
=q_\ell^{\ii u(r-s)}E_\ell(r,s).
\end{equation}
In particular, each vacuum projection
$E_\ell(0,0)$ is modularly invariant.

\subsection{Occupation-number subspaces and their weights}
\label{app:occupation:weights}

Define the simultaneous vacuum projection
$\Pi_0:=\bigwedge_{\ell\in\mathcal I}
E_\ell(0,0)$, the strong limit of the finite
products of vacuum projections.
By Lemma~\ref{lem:finite-factor} and normality,
$\Tr(\mu\Pi_0)=c>0$.

For each finite-support occupation sequence
$\mathbf n\in\N_0^{(\mathcal I)}$, define
\begin{equation}\label{eq:occupation-isometries}
T_{\mathbf n}:=
\left(\prod_{\ell:n_\ell>0}
E_\ell(n_\ell,0)\right)\Pi_0,
\qquad
\Pi_{\mathbf n}:=T_{\mathbf n}T_{\mathbf n}^*.
\end{equation}
The matrix-unit relations give
\begin{equation}\label{eq:occupation-orthogonality}
T_{\mathbf n}^*T_{\mathbf m}
=\delta_{\mathbf n\mathbf m}\Pi_0.
\end{equation}
Hence the projections $\Pi_{\mathbf n}$ are mutually orthogonal. We call $\Ran\Pi_{\mathbf n}$ the \emph{occupation-number sector} associated with $\mathbf n$. Each $T_{\mathbf n}$ maps the vacuum subspace isometrically onto this subspace.

By \eqref{eq:occupation-modular},
\[
\alpha_u(T_{\mathbf n})
=(q^{\mathbf n})^{\ii u}T_{\mathbf n},
\qquad
q^{\mathbf n}:=\prod_\ell q_\ell^{n_\ell}.
\]
If a bounded operator $X$ satisfies
$\alpha_u(X)=\lambda^{\ii u}X$ for all
$u\in\R$, then
\begin{equation}\label{eq:modular-eigenoperator}
\mu X=\lambda X\mu.
\end{equation}
Indeed, writing $\mu e_r=u_re_r$, the modular
identity implies $X_{rs}=0$ unless
$u_r=\lambda u_s$, which proves
\eqref{eq:modular-eigenoperator}.

Applying this to $T_{\mathbf n}$ gives
$\mu T_{\mathbf n}
=q^{\mathbf n}T_{\mathbf n}\mu$.
Using \eqref{eq:occupation-orthogonality},
we obtain
\[
\begin{aligned}
\Tr(\mu\Pi_{\mathbf n})
&=\Tr(T_{\mathbf n}^*\mu T_{\mathbf n})\\
&=q^{\mathbf n}\Tr(\mu\Pi_0)
=cq^{\mathbf n},
\end{aligned}
\]
which proves \eqref{eq:occupation-weights}.

\subsection{The Fock identification and density factorization}
\label{app:occupation:factorization}
Set $\cH_K:=\Pi_0\cH_U$.
For the standard finite-support occupation basis
$\{|\mathbf n\rangle\}$ of
$\cH_Q=\cF(\ell^2(\mathcal I;\C))$, define
\begin{equation}\label{eq:occupation-unitary}
\mathcal U(|\mathbf n\rangle\otimes v)
:=T_{\mathbf n}v,
\qquad v\in\cH_K.
\end{equation}

The orthogonality relation
\eqref{eq:occupation-orthogonality} shows that this map
preserves inner products on finite linear combinations
of occupation basis vectors. Since those vectors are
dense in the Fock tensor product, it extends to an
isometry

$\mathcal U:\cH_Q\otimes\cH_K\to\cH_U$,
whose range projection is
$\sum_{\mathbf n}\Pi_{\mathbf n}$.
Equation~\eqref{eq:exhaustion} and faithfulness
of $\mu$ show that $\mathcal U$ is onto.

It remains to identify the density under
this unitary. From
\eqref{eq:modular-eigenoperator} and
\eqref{eq:occupation-orthogonality},
\begin{equation}\label{eq:occupation-density-blocks}
T_{\mathbf n}^*\mu T_{\mathbf m}
=\delta_{\mathbf n\mathbf m}
q^{\mathbf n}\Pi_0\mu\Pi_0.
\end{equation}
Consequently,
\[
\mathcal U^*\mu\mathcal U
=
\left(\sum_{\mathbf n}q^{\mathbf n}
|\mathbf n\rangle\langle\mathbf n|\right)
\otimes\left(\Pi_0\mu\Pi_0|_{\cH_K}\right).
\]
Since $\Tr(\mu\Pi_0)=c$, define
$\nu:=c^{-1}\Pi_0\mu\Pi_0|_{\cH_K}$.
The product formula for $c$ then gives
\[
\mathcal U^*\mu\mathcal U
=\tau_Q\otimes\nu,
\qquad
\tau_Q:=c\sum_{\mathbf n}q^{\mathbf n}
|\mathbf n\rangle\langle\mathbf n|.
\]
Both factors are faithful normal states.
This completes the density factorization
used in Proposition~\ref{prop:fock}.

\section{The thermal sandwich identity}
\label{app:sandwich}

We prove \eqref{eq:sandwich}, first for one mode
and then for countably many modes.

For $0<q<1$, the canonical purification of
$\tau(q)$ is
\[
|\sqrt\tau\rangle
:=\sqrt{1-q}\sum_{r\ge0}q^{r/2}|r,r\rangle.
\]
Let $a_1,a_2$ be the annihilation operators
of the two modes. The operators
\[
b_1:=\frac{a_1-\sqrt q\,a_2^*}{\sqrt{1-q}},
\qquad
b_2:=\frac{a_2-\sqrt q\,a_1^*}{\sqrt{1-q}}
\]
satisfy the canonical commutation relations
and annihilate $|\sqrt\tau\rangle$.
Since this vector has finite number expectation,
these identities hold on the corresponding
operator domains. The associated symplectic
transformation is unitarily implementable
\cite[Theorem~6]{Derezinski}, so
$|\sqrt\tau\rangle$ is a Gaussian vector.

Its covariance matrix is
\begin{equation}\label{eq:purification-covariance}
\begin{pmatrix}
vI_2&c_qZ\\
c_qZ&vI_2
\end{pmatrix},
\qquad
v:=\frac{1+q}{1-q},\quad
c_q:=\frac{2\sqrt q}{1-q},\quad
Z:=\begin{pmatrix}1&0\\0&-1\end{pmatrix}.
\end{equation}
Indeed, the occupation-number expansion gives
$\langle a_j^*a_j\rangle=q/(1-q)$ and
$\langle a_1a_2\rangle=\sqrt q/(1-q)$,
while $\langle a_j^2\rangle
=\langle a_1^*a_2\rangle=0$.

For bounded operators $X,Y$, expansion in the
number basis yields
\begin{equation}\label{eq:vectorization}
\Tr(\tau^{1/2}X\tau^{1/2}Y)
=\langle\sqrt\tau|
X^{\mathsf T}\otimes Y
|\sqrt\tau\rangle.
\end{equation}
In this basis, $Q^{\mathsf T}=Q$ and
$P^{\mathsf T}=-P$, so
$W(a)^{\mathsf T}=W(Za)$.
Taking $X=W(a)$ and $Y=W(f)$ in
\eqref{eq:vectorization}, the covariance
\eqref{eq:purification-covariance} gives
\[
\begin{aligned}
\Tr[\tau^{1/2}W(a)\tau^{1/2}W(f)]
&=\exp\left[
-\tfrac14v|a|^2
-\tfrac12c_q a^{\mathsf T}f
-\tfrac14v|f|^2
\right].
\end{aligned}
\]
Here the mixed term follows from
$(Za)^{\mathsf T}c_qZf=c_q a^{\mathsf T}f$.

Taking tensor products proves
\eqref{eq:sandwich} when $a,f\in\mathcal Y$
have finite support. For general
$a\in\mathcal Y$, choose finite-support
truncations $a_N\to a$. Strong continuity
of the Fock Weyl operators and the
Hilbert--Schmidt inequality give
\[
\begin{aligned}
&\norm{\tau_Q^{1/2}
(W_Q(a_N)-W_Q(a))\tau_Q^{1/2}}_1\\
&\quad\le
\norm{\tau_Q^{1/2}}_{\HS}
\norm{(W_Q(a_N)-W_Q(a))
\tau_Q^{1/2}}_{\HS}
\longrightarrow0.
\end{aligned}
\]
The last convergence follows by finite-rank
approximation of $\tau_Q^{1/2}$.
Since $\mathsf V$ and $\mathsf C$ are bounded,
the quadratic terms converge as well.
This proves \eqref{eq:sandwich}.

\bibliographystyle{unsrt}
\bibliography{references}

@misc{PetzCCR,
  author = {Petz, D{\'e}nes},
  title = {The Algebra of the Canonical Commutation Relation},
  year = {2011},
  note = {Lecture notes, November 21, 2011},
  url = {https://math.bme.hu/~petz/CCR.pdf}
}

@article{Kac,
  author = {Kac, Mark},
  title = {On a Characterization of the Normal Distribution},
  journal = {American Journal of Mathematics},
  volume = {61},
  number = {3},
  pages = {726--728},
  year = {1939},
  doi = {10.2307/2371328}
}

@article{Bernstein,
  author = {Bernstein, S. N.},
  title = {On a Property Which Characterizes {Gauss}'s Law},
  journal = {Trudy Leningradskogo Politekhnicheskogo Instituta},
  pages = {21--22},
  year = {1941},
  note = {In Russian; title rendered in English translation}
}

@article{JPP,
  author = {Jen{\v{c}}ov{\'a}, Anna and Petz, D{\'e}nes and Pitrik, J{\'o}zsef},
  title = {Markov Triplets on {CCR}-Algebras},
  journal = {Acta Scientiarum Mathematicarum (Szeged)},
  volume = {76},
  number = {1--2},
  pages = {111--134},
  year = {2010},
  eprint = {0812.3724},
  archivePrefix = {arXiv},
  url = {https://acta.bibl.u-szeged.hu/16340/}
}

@article{HJPW,
  author = {Hayden, Patrick and Jozsa, Richard and Petz, D{\'e}nes and Winter, Andreas},
  title = {Structure of States Which Satisfy Strong Subadditivity of Quantum Entropy with Equality},
  journal = {Communications in Mathematical Physics},
  volume = {246},
  pages = {359--374},
  year = {2004},
  doi = {10.1007/s00220-004-1049-z},
  eprint = {quant-ph/0304007},
  archivePrefix = {arXiv}
}

@article{JP,
  author = {Jen{\v{c}}ov{\'a}, Anna and Petz, D{\'e}nes},
  title = {Sufficiency in Quantum Statistical Inference},
  journal = {Communications in Mathematical Physics},
  volume = {263},
  pages = {259--276},
  year = {2006},
  doi = {10.1007/s00220-005-1510-7},
  eprint = {math-ph/0412093},
  archivePrefix = {arXiv}
}

@article{DePalmaTrevisan,
  author = {De Palma, Giacomo and Trevisan, Dario},
  title = {The Conditional Entropy Power Inequality for Bosonic Quantum Systems},
  journal = {Communications in Mathematical Physics},
  volume = {360},
  pages = {639--662},
  year = {2018},
  doi = {10.1007/s00220-017-3082-8}
}

@article{Weedbrook,
  author = {Weedbrook, Christian and Pirandola, Stefano and Garc{\'i}a-Patr{\'o}n, Ra{\'u}l and Cerf, Nicolas J. and Ralph, Timothy C. and Shapiro, Jeffrey H. and Lloyd, Seth},
  title = {Gaussian Quantum Information},
  journal = {Reviews of Modern Physics},
  volume = {84},
  number = {2},
  pages = {621--669},
  year = {2012},
  doi = {10.1103/RevModPhys.84.621}
}

@book{BratteliRobinson,
  author = {Bratteli, Ola and Robinson, Derek W.},
  title = {Operator Algebras and Quantum Statistical Mechanics II: Equilibrium States, Models in Quantum Statistical Mechanics},
  series = {Theoretical and Mathematical Physics},
  edition = {1},
  publisher = {Springer},
  address = {Berlin, Heidelberg},
  year = {1981},
  doi = {10.1007/978-3-662-09089-3}
}

@article{GengNair,
  author = {Geng, Yanlin and Nair, Chandra},
  title = {The Capacity Region of the Two-Receiver {Gaussian} Vector Broadcast Channel with Private and Common Messages},
  journal = {IEEE Transactions on Information Theory},
  volume = {60},
  number = {4},
  pages = {2087--2104},
  year = {2014},
  doi = {10.1109/TIT.2014.2304457}
}

@article{AnantharamJogNair,
  author = {Anantharam, Venkat and Jog, Varun and Nair, Chandra},
  title = {Unifying the {Brascamp--Lieb} Inequality and the Entropy Power Inequality},
  journal = {IEEE Transactions on Information Theory},
  volume = {68},
  number = {12},
  pages = {7665--7684},
  year = {2022},
  doi = {10.1109/TIT.2022.3192913}
}

@article{ElGamalGohariNair,
  author = {{El Gamal}, Abbas and Gohari, Amin and Nair, Chandra},
  title = {A Strengthened Cutset Upper Bound on the Capacity of the Relay Channel and Applications},
  journal = {IEEE Transactions on Information Theory},
  volume = {68},
  number = {8},
  pages = {5013--5043},
  year = {2022},
  doi = {10.1109/TIT.2022.3163647}
}

@article{Springer,
  author = {Springer, S. C. and Lee, Jinhyoung and Bellini, Marco and Kim, M. S.},
  title = {Conditions for Factorizable Output from a Beam Splitter},
  journal = {Physical Review A},
  volume = {79},
  number = {6},
  pages = {062303},
  year = {2009},
  doi = {10.1103/PhysRevA.79.062303}
}

@article{Cuesta,
  author = {Cuesta, Javier},
  title = {A Stable Quantum {Darmois--Skitovich} Theorem},
  journal = {Journal of Mathematical Physics},
  volume = {61},
  pages = {022201},
  year = {2020},
  doi = {10.1063/1.5122955}
}

@incollection{Derezinski,
  author = {Derezi{\'n}ski, Jan},
  title = {Introduction to Representations of the Canonical Commutation and Anticommutation Relations},
  booktitle = {Large Coulomb Systems},
  editor = {Derezi{\'n}ski, Jan and Siedentop, Heinz},
  series = {Lecture Notes in Physics},
  volume = {695},
  pages = {63--143},
  publisher = {Springer},
  address = {Berlin, Heidelberg},
  year = {2006},
  doi = {10.1007/3-540-32579-4_3}
}

@article{Takesaki,
  author = {Takesaki, Masamichi},
  title = {Conditional Expectations in von Neumann Algebras},
  journal = {Journal of Functional Analysis},
  volume = {9},
  number = {3},
  pages = {306--321},
  year = {1972},
  doi = {10.1016/0022-1236(72)90004-3}
}

@book{Krengel,
  author = {Krengel, Ulrich},
  title = {Ergodic Theorems},
  series = {De Gruyter Studies in Mathematics},
  volume = {6},
  publisher = {Walter de Gruyter},
  address = {Berlin},
  year = {1985},
  doi = {10.1515/9783110844641}
}

@book{Kannappan,
  author = {Kannappan, Palaniappan},
  title = {Functional Equations and Inequalities with Applications},
  series = {Springer Monographs in Mathematics},
  publisher = {Springer},
  address = {New York},
  year = {2009},
  doi = {10.1007/978-0-387-89492-8}
}
\end{document}